\documentclass[msc, anglais, dvi, index, reqno]{theseUL}

\newcommand{\ee}[1]{\mbox{${} \times 10^{#1}$}}\newcommand{\msun}{\mbox{$M_\odot$}}

\usepackage{astron}
\usepackage{graphicx}
\usepackage{deluxetable}
\makeindex

\usepackage{subfigure}
\begin{document}

%
%
%


\def\jnl@style{\it}
\def\aaref@jnl#1{{\jnl@style#1}}

\def\aaref@jnl#1{{\jnl@style#1}}

\def\aj{\aaref@jnl{AJ}}                   
\def\araa{\aaref@jnl{ARA\&A}}             
\def\apj{\aaref@jnl{ApJ}}                 
\def\apjl{\aaref@jnl{ApJ}}                
\def\apjs{\aaref@jnl{ApJS}}               
\def\ao{\aaref@jnl{Appl.~Opt.}}           
\def\apss{\aaref@jnl{Ap\&SS}}             
\def\aap{\aaref@jnl{A\&A}}                
\def\aapr{\aaref@jnl{A\&A~Rev.}}          
\def\aaps{\aaref@jnl{A\&AS}}              
\def\azh{\aaref@jnl{AZh}}                 
\def\baas{\aaref@jnl{BAAS}}               
\def\jrasc{\aaref@jnl{JRASC}}             
\def\memras{\aaref@jnl{MmRAS}}            
\def\mnras{\aaref@jnl{MNRAS}}             
\def\pra{\aaref@jnl{Phys.~Rev.~A}}        
\def\prb{\aaref@jnl{Phys.~Rev.~B}}        
\def\prc{\aaref@jnl{Phys.~Rev.~C}}        
\def\prd{\aaref@jnl{Phys.~Rev.~D}}        
\def\pre{\aaref@jnl{Phys.~Rev.~E}}        
\def\prl{\aaref@jnl{Phys.~Rev.~Lett.}}    
\def\pasp{\aaref@jnl{PASP}}               
\def\pasj{\aaref@jnl{PASJ}}               
\def\qjras{\aaref@jnl{QJRAS}}             
\def\skytel{\aaref@jnl{S\&T}}             
\def\solphys{\aaref@jnl{Sol.~Phys.}}      
\def\sovast{\aaref@jnl{Soviet~Ast.}}      
\def\ssr{\aaref@jnl{Space~Sci.~Rev.}}     
\def\zap{\aaref@jnl{ZAp}}                 
\def\nat{\aaref@jnl{Nature}}              
\def\iaucirc{\aaref@jnl{IAU~Circ.}}       
\def\aplett{\aaref@jnl{Astrophys.~Lett.}} 
\def\apspr{\aaref@jnl{Astrophys.~Space~Phys.~Res.}}
\def\bain{\aaref@jnl{Bull.~Astron.~Inst.~Netherlands}} 
\def\fcp{\aaref@jnl{Fund.~Cosmic~Phys.}}  
\def\gca{\aaref@jnl{Geochim.~Cosmochim.~Acta}}   
\def\grl{\aaref@jnl{Geophys.~Res.~Lett.}} 
\def\jcp{\aaref@jnl{J.~Chem.~Phys.}}      
\def\jgr{\aaref@jnl{J.~Geophys.~Res.}}    
\def\jqsrt{\aaref@jnl{J.~Quant.~Spec.~Radiat.~Transf.}}
\def\memsai{\aaref@jnl{Mem.~Soc.~Astron.~Italiana}}
\def\nphysa{\aaref@jnl{Nucl.~Phys.~A}}   
\def\physrep{\aaref@jnl{Phys.~Rep.}}   
\def\physscr{\aaref@jnl{Phys.~Scr}}   
\def\planss{\aaref@jnl{Planet.~Space~Sci.}}   
\def\procspie{\aaref@jnl{Proc.~SPIE}}   

\let\astap=\aap
\let\apjlett=\apjl
\let\apjsupp=\apjs
\let\applopt=\ao

\PrenomNomMaj{WILLIAM BRITO}
\PrenomNomMin{William Brito}
\titre{The Fate of Dwarf Galaxies in Clusters and 
the Origin of Intracluster Stars}

\programme{en physique}
\grade{Ma\^itre \`es Sciences (M.Sc.)}
\faculte{D\'EPARTEMENT DE PHYSIQUE, DE G\'ENIE PHYSIQUE ET D'OPTIQUE
FACULT\'E DES SCIENCES ET DE G\'ENIE}
\annee{2009}
\maketitle

\chapter*{R\'esum\'e}

Ce m\'emoire r\'esume quelques concepts importants en cosmologie et pr\'esente l'\'etude faite par l'auteur sur l'origine de la lumi\`ere intra-amas.

Pour la r\'ealisation de ce projet, l'auteur a tout d'abord recherch\'e dans la litt\'erature les param\`etres \`a utiliser pour des simulations en langage FORTRAN dont les algorithmes de base sont, dans la premi\`ere partie du projet, particule-particule et, dans la seconde, particule-particule/particule-maille.  L'auteur a \'egalement modifi\'e des codes IDL et UNIX.  Enfin, le projet n\'ecessita des centaines de simulations d'amas isol\'es dont les r\'esultats ont \'et\'e analys\'es en collaboration avec les membres du groupe de recherche et soumis pour publication (Barai, Brito \& Martel 2009).

Les r\'esultats principaux des simulations d\'ecrites dans ce document sont: 1) la destruction des galaxies naines par des fusions domine sur la destruction par des mar\'ees, et 2) la destruction des galaxies par des mar\'ees est suffisante pour expliquer la lumi\`ere intra-amas observ\'ee.

Finalement, les r\'esultats d'amas isol\'es ont \'et\'e g\'en\'eralis\'es \`a une r\'egion significative de l'Univers.  Ainsi, l'auteur a contribu\'e \`a la mise en oeuvre d'une simulation particule-particule/particule-maille et \`a l'analyse commune des r\'esultats obtenus \`a ce jour.  Les r\'esultats reproduisent la fonction de luminosit\'e de Schechter, et sugg\`erent que l'approche utilis\'ee est valide et que les r\'esultats sont robustes.

\chapter*{Abstract}

This thesis presents a review of related important concepts in cosmology followed by details of the author's role in a research project on the origin of intracluster light.   

The author's role in the development of the simulations varied from searching parameters in the literature, through writing and modifying code in IDL, FORTRAN, and UNIX to carrying out hundreds of simulations using the particle-particle algorithm described in this thesis, as well as partaking in joint analysis of the simulation results.  Part of this work in the isolated cluster simulations has been submitted for publication (Barai, Brito \& Martel 2009).

The main results of the simulations described in this thesis are: 1) destruction of dwarf galaxies by mergers dominates destruction by tides, and 2) destruction of galaxies by tides is sufficient to explain the observed intracluster light.  These results support the accepted explanation for the origin of the intracluster stars. 

In an ongoing, second stage of the simulation, which extends the isolated cluster results to a cosmologically significant region of the Universe, the author similarly assists in the implementation of a particle-particle/particle-mesh simulation and the joint analysis of the results to-date.  The results are as per the Schechter luminosity function, and suggest the approach used is valid and the results obtained robust.

\chapter*{Avant-propos}
The list of people to thank is endless, as usual.  Thanks to everyone in my family, to be sure, for their patience, but within the university, I thank particularly Laurent Drissen for encouraging me to proceed towards the master's degree when he did\,\textemdash\,surprising what a word at the right time can do!    Thanks to Serge Pineault for his understanding and encouragement along the way, especially when I had recently returned to the study of science.  Thanks to other graduate students, such as Simon Cantin, Jean-Fran\c cois Robitaille, and V\'eronique Petit, for answering my many questions, sometimes repeatedly!  Thanks to my friend Andrea Clark for copy editing so many mistakes out of this text\,\textemdash\,at such short notice!  Thanks last (certainly not least!) to Paramita Barai and my Director Hugo Martel for their guidance, patience, and understanding throughout my time studying with them.  While the above individuals (and many others) have helped a great deal, any error surviving to the printing stage is exclusively my own.

\begin{dedicace}

To my mom, Wilma Webster, and my dad, Urbano Brito, who always encourage  me to continue learning;
to uncle Jim Webster, and to my siblings Cendy, Moy, and Steven who are also ever in my heart and mind.
\end{dedicace}

\clearpage

\begin{epigraphe}

\begin{verbatim}



Release me for a little while
And let me go
Out in the twilight
Where over red fires of sunset
The thin moon gleams
There let me pass
Let mortal shadows drop away
And leave the spinning earth
Behind me
In the weightlessness of outer space
I would adventure
Towards far galaxies

--Andromeda Webster
 Flashes from space.
\end{verbatim}

\begin{verbatim}





















Lo!  That some we loved, the loveliest
and the best of all time's vintage pressed, 
have drunk their cup a round or two before
and crept silently to rest!

--Omar Khayam
 The Rubaiyat
\end{verbatim}
\end{epigraphe}

\tableofcontents
\listoftables
\listoffigures 
\corps

\chapter{Introduction}

\begin{figure} [!htp]
	\centering
	\begin {center}
	\hskip 100pt
		\includegraphics[bb=20 20 675 660,clip=true,viewport=05 89 924 700,
		scale=.75, angle=0]{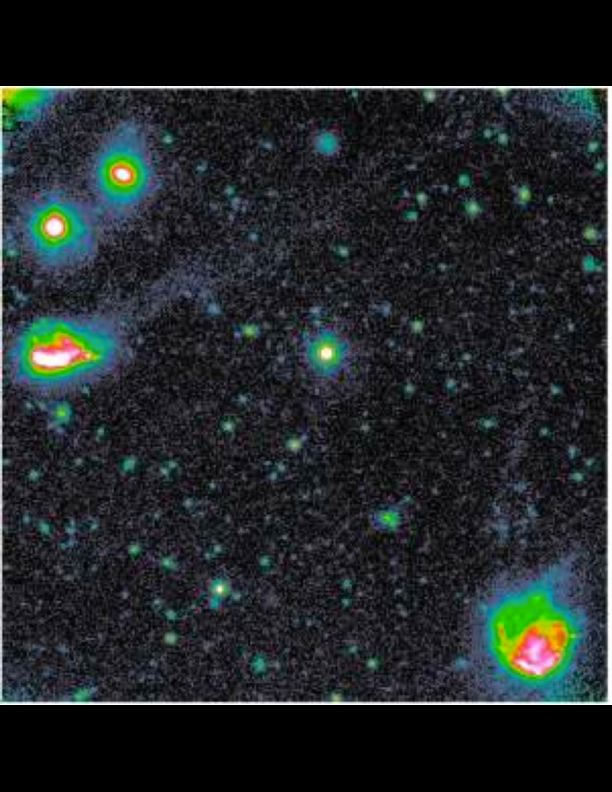}
	\vskip 25pt \caption[Trails of gas in a nearby cluster.]{\textbf{Trails of gas in a nearby cluster.}  \emph{Trails of gas and stars appear to spring from two galaxies (at the left and in the bottom right corner) in the Abell 1367 cluster.  The trails are probably due to a tidal event 50 million years ago.  Image credit: Peppo Gavazzi, Universita di Milano Bicocca/MPIA.}}
	\label{abell1367}
	\end {center}
\end{figure}

Since it was first observed, the origin and nature of the intracluster (IC) light, or diffuse light, has been a mystery to astronomers.  With the advent of more powerful telescopy the existence of IC stars became clear (Fig. \ref{abell1367}) and the question became ``Where do these stars originate?"  These stars could be part of cluster-member galaxies; they could have formed in-situ;  or they could have been stripped from member galaxies by various processes.
Dwarf galaxies (DGs) are the most numerous galaxies; they are low mass ($10^7\,\textendash\,10^9\msun$) galaxies having an absolute magnitude fainter than $\rm{M_B} \approx{-16 }$ to $-18$ mag \cite{groeb,bothunElsb}.  Contemporary opinion ascribes the origin of IC stars to, for example, the tidal destruction of DGs during the evolution of the cluster.  Clusters are large groups containing up to thousands of galaxies.  DGs can be destroyed in various ways during the evolution of the cluster.  Understanding the nature of IC light is important as, for example, reproduction of the observed proportions (and other related parameters) of IC light can be used to refine cosmological models.  The goal of this research project is to compare the various modes of dwarf galaxy destruction and ascertain if dwarf galaxy destruction by tides is sufficient to explain the observed IC light.

As part of a team I have simulated and examined the relationships between six possible outcomes of dwarf galaxy destruction during the evolution of isolated clusters.  Subsequently, with this team, I have analyzed the results to date of an extension of the isolated cluster results to a cosmologically significant region of the Universe.  The analysis, and the simulations I carried out as member of the team in this project use known parameters such as the luminosity function (LF), various cluster halo mass profiles, and the cosmological parameters to simulate, test, and corroborate the method used and the results obtained.

Typical simulations exploring the evolution of clusters devote much time and processing power to a single simulation.  Shifting attention from sheer processing power to clever programming, however, can allow for statistical exploration of the relationships, in a significant region of the Universe, between different modes of DG destruction in clusters.  DG destruction has been widely touted as an explanation for the observed IC light, which contributes  up to 25\% to cluster luminosities.  The destruction of DGs (by tides) would produce IC stars in the IC medium; it is these stars that are the purported source of the observed IC light.
     
     Heuristic modeling of the DG interactions allows prediction of the average DG behavior with an economy of time and processing power.  Rather than unnecessarily flex processing power ``muscles"\kern -3pt, subgrid physics are implemented in this project and a small number of particles ($ \sim1\ 000)$ is used to represent a typical cluster. Results are consistent with accepted theories of the IC light being largely produced by tidal disruption of DGs. 

Alternative theories as to the origin of the IC stars include for example, in situ formation, with the notion that the IC stars originate at least partly from DGs that have had the outermost material stripped by a process known as `galaxy harassment' \cite{richstone76,moorbook}, or another phenomenon termed tidal stirring, where tidal shocks strip the DGs \cite{mayer01}.

In this research project six possible outcomes are identified from interaction between a dwarf galaxy and the rest of the cluster: 
(1) The galaxy merges with another; (2) The galaxy is destroyed by the tidal field of a larger galaxy, but the fragments accrete to that larger galaxy; (3) The galaxy is destroyed by tides of a larger galaxy, and the fragments are dispersed in the IC medium; (4) The galaxy is destroyed by the tidal field of the background halo; (5) The galaxy survives to the present (i.e. it is not destroyed by any process); and (6) The galaxy is ejected from the cluster.  By analyzing these interactions it is shown that the average behavior of DGs is consistent with the destruction of DGs as an explanation for the origin of IC stars when one takes into account the various results of interaction.  That is, IC stars come from a variety of origins, which for the most part arise from DG interactions in the cluster.  

The results from this part of the project, obtained from simulations of a single idealized cluster are extended in a second, ongoing half of the project that simulates a cosmologically significant volume of space, i.e. a volume containing numerous clusters.

Study of the IC light is timely, as recent discoveries of IC stars in the Virgo cluster might result in an opportunity to learn about  details of cluster origins \cite{arnaMag1}. Additionally, understanding properties of clusters, such as the luminosity function, number density function, and formation epoch is worthwhile as these are important tools in testing cosmological models and parameters.

 \section{Dwarf galaxies}

Galaxies in general can be thought of as the building blocks of the Universe; DGs can be thought of as the fundamental units of mass.  DGs are the galaxies that have the largest dark matter (DM) component \cite{bothunElsb} and they are the most numerous.  Nonetheless, most studies conclude the DG contribution to the total mass density is smaller than that of giant galaxies, stemming from the untested assumption that the mass-to-light ratio ($\rm{M/L}$) of galaxies does not increase for smaller and smaller galaxies \cite{bothunElsb}.  The simulations presented here include only gravity, with no other physical process (gas and DM behave the same way under gravity).  When necessary, the IC light is calculated using a $\rm{M/L}$.

 DGs, defined in the previous section as low mass ($10^7\,\textendash\,10^9\msun$) galaxies having an absolute magnitude fainter than $\rm{M_B} \approx{-16 }$ or ${-18}$ mag, have low surface brightness and low metallicity \cite{groeb}.  In the simulations presented herein,  as in the observed Universe, DGs are the most numerous galaxies occurring.  DGs are thought to comprise approximately 80\% of the Local Volume galaxy population ($D \leq10$ Mpc), and may have a space density $\sim$ 40 times that of bright galaxies in the Universe \cite{dgobs}.  There exists, however, a deficiency in the number of observed low luminosity DGs (discrepancy more than one order of magnitude) as compared with the large number of theoretically predicted low mass DM haloes \cite{virgolf,dgdynG}.   This is recognized as a problem for cold DM theory; the likely solution involves energy feedback from stellar evolution.

 DGs belong typically to one of three main morphological types: (1) dwarf irregulars (dIrr),  (2) dwarf spheroidals (dSph), and (3) dwarf ellipticals (dE). These types of galaxies have similar scaling, but some dwarf irregulars have bluer colors and more disturbed morphologies; they are called blue compact dwarfs.  There exist other, rare galaxies that do not fit into this classification, such as huge low surface brightness giants like Malin 1 \cite{flumtrent}.
 
 There is evidence that dEs are further subdivided into two classes, compact ones and diffuse ones.  The only compact dE in the Local Group is M32, which lies very close to the Milky Way's giant companion, M31.  Compact dE galaxies that are not close companions of any giant galaxy have been found in nearby galaxy clusters.  The radial surface brightness profiles of these galaxies are reasonably well fitted by a $R^{1/4}$ law, corresponding to their elliptical nature; this is called the de Vaucouleurs law, as per Binney \& Merrifield \cite*{bm98}, hereafter BM98.  By contrast, the radial surface-brightness profiles of the other class of dE galaxies, the diffuse ones, are best fitted by an exponential
$e^ {-R/{R_{\rm d}}}$,

Bothun et al. (1991) mention that there are at least two other profile types.  One is that some dwarfs have very flat cores and then steep fall-offs; in general these objects are at the faint end of the distribution in
central surface brightness, and consequently they are the most difficult to discover. Another is that some compact dwarfs have power-law surface brightness profiles.  If sufficiently faint or small, these galaxies are difficult to distinguish from star clusters.

The DGs in the simulations of this project are assumed to be spherical (an idealization). When calculating galaxy-pair interactions, however, a galaxy is herein considered a bound virialized system with an internal kinetic energy and a potential energy.  These energies are included in the total energy of the interacting pair of galaxies (Barai et al. 2009).

\section{Dark matter}

The as-of-yet unobserved dark matter (DM) is a pressureless fluid thought to be responsible for the growth of cosmological perturbations through gravitational instability; it is expected to be more abundant in extensive haloes that stretch to 100\,\textendash\,200 kpc from the center of galaxies \cite{verdcosm}. 

The single systems with the largest proportion of DM are believed to be faint DGs.  DGs in general have a correspondingly high ratio of dark-to-luminous mass, with $\rm{M/L}$ as high as that of galaxy groups and poor clusters.  DGs are increasingly DM-dominated at fainter magnitudes.   Among the faintest dwarfs in the Local Group are DGs such as Draco and Ursa Minor with $\rm{M/L}\approx100$ within their core-fitting radii and total $\rm{M/L}$ that are probably much larger than this \cite{flumtrent}\footnotemark \footnotetext{The high DM content of  these faint DGs suggests a link between their mass functions and the power spectrum of primordial fluctuations of small scales (Trentham \& Tully 2002).}.%

\begin{figure} [!htp]
	\centering
	\begin {center}
		
		\includegraphics[bb=20 20 675 660,clip=true,viewport=05 69 924 722,
                  scale=.75, angle=0]{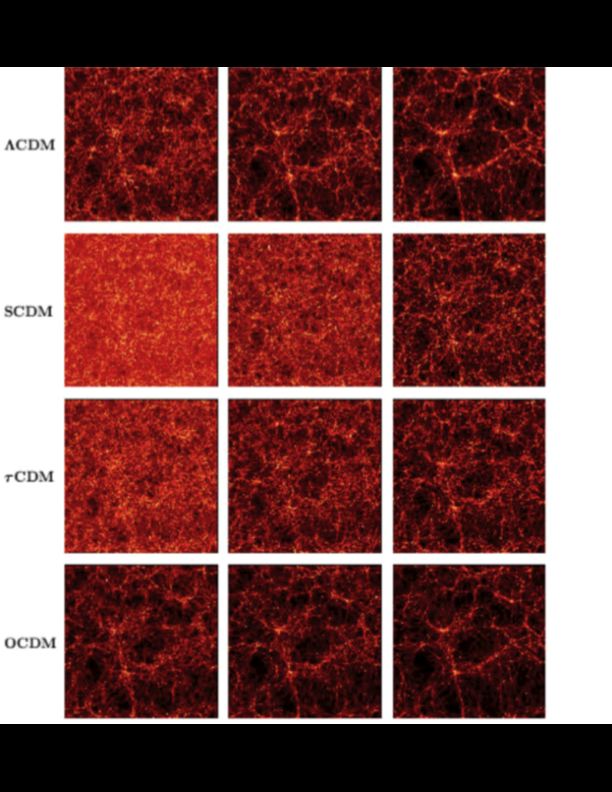}

	\vskip 20pt \caption[Predictions of various cold dark matter models]{\textbf{Predictions of various  cold DM models.}  \emph{States for three redshifts are shown.  $\Lambda$CDM is the concordance model; it includes a cosmological constant.  The standard model (SCDM) does not assume a cosmological constant.  The tilted model ($\tau$CDM) takes into account possible pre-inflationary inhomogeneities.  The open model (OCDM) produces results consistent with the observed Universe but does not assume the Universe is flat (flatness is more consistent with inflation).  Image credit: (GIF)~Colberg~(1997).}}
	\label{1diffmpreds}
	\end {center}
\end{figure}

 It is thought today that of the total matter in the Universe, an overwhelming 95\% of the mass in galaxies and galaxy clusters is made of an unknown DM component, the vast majority of which is non-relativistic (cold).  Evidence for the existence of DM comes from measurements of the rotation curves of galaxies, measurements of gravitational lensing and  from the existence of hot gas in clusters \cite{freesdark} as well as from methods of estimating cluster masses on the basis of the temperature of IC gas \cite{rosevol}.  Evidence also comes from measurements including cosmic microwave background temperature fluctuations, redshift surveys, gravitational-lensing shear effects, and distribution of absorbing neutral clouds along different lines of sight \cite{pryormods}.   The \textit{concordance} cosmological model, $\Lambda$CDM (where $\Lambda$ is the cosmological constant and CDM stands for cold DM), is the simplest model resulting in an interpretation consistent with the observed Universe.  Figure~\ref{1diffmpreds} shows the results of various simulations of which the $\Lambda$CDM model produces voids and filaments closer to observations of the actual Universe.
 
The total density parameter is $\Omega_0=0.27$. Baryons account for $\Omega_{\rm b}=0.04$. Hence, the non-baryonic (exotic) part accounts for $0.23/0.27=85\%$ of the mass in the Universe, and this component is what is normally referred to as DM.  Notice, however that some of the baryonic matter is unobserved.  This component is called baryonic DM.  The most likely candidates for dark baryons are known as massive compact halo objects  (MACHOs), which include faint stars or stellar remnants.  The best current limit on the MACHO abundance is ²20\% of the dark halo mass.  Interestingly, the appearance of baryons varies with scale, occurring mostly in stars below a galaxy mass scale of $M^*$ ($M^*$ is the mass of the Galaxy, the Milky Way) and mostly in hot gas for systems much more massive than $M^*$ (i.e. galaxy groups and clusters), while for DM there is no difference in properties between galaxy, group, and cluster scales \cite{silkgform}.  Further progress on DM structure will come as progress is made in understanding when and how galaxies formed  \cite{silkgform}.  Currently, it is thought that a realistic distribution of DM haloes might be produced by models including a process involving feedback augmented by, for example, hyper novae (a very energetic supernova thought to result from an extreme core-collapse scenario), or a top-heavy initial mass function (the mass dependence of the number of stars that form per mass interval per unit volume).  Models resulting in such realistic distributions can involve for example incorporating star and active galactic-nuclei (AGN) feedback; using massive gas outflows can effectively weaken the DM gravity in the central cusp of the dense baryonic core, which forms by gas dissipation \cite{silkgform}.

The presented simulations assume an isotropic distribution of the galaxies inside the cluster; this is an idealization, but little is yet known of the internal structure of the DM as explained in the foregoing.  Furthermore, the positions of the galaxies herein are chosen using realistic guidelines consistent with the literature.  

\section{Luminosity functions}

\begin{figure} [htp]
	\centering
	\begin {center}
		
		\includegraphics[bb=00 00 605 640,clip=true,viewport=12 9 600 792,
                  scale=.7, angle=0]{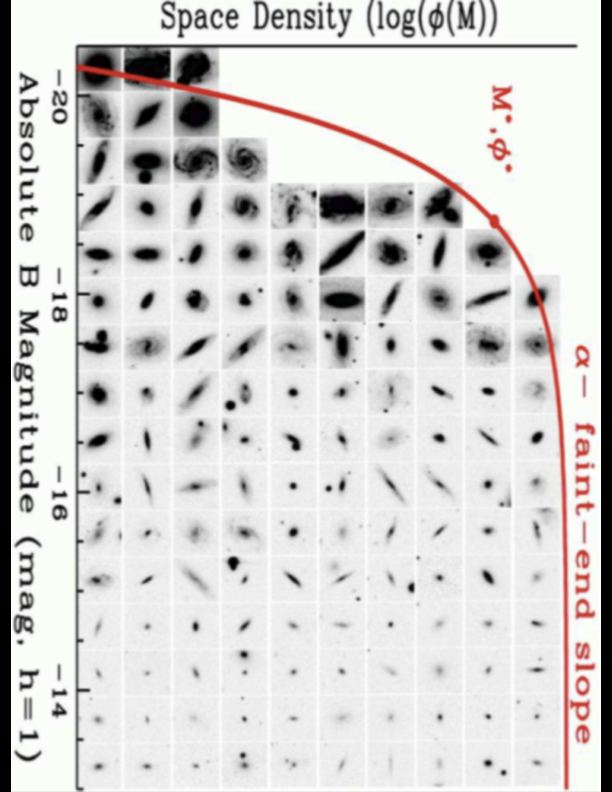}

	\vskip 25pt \caption[The global luminosity function]{\textbf{The global luminosity function.}  \emph{Shown here is an example LF for a nearby volume-limited sample and images of the actual galaxies contributing to the LF.  Image borrowed from Driver (2004).}}
	\label{globlum}
	\end {center}
\end{figure}

\begin{figure} [!htp]
	\centering
	\begin {center}
		
		\includegraphics[bb=00 00 605 640,clip=true,viewport=12 49 600 742,
                  scale=.77, angle=0]{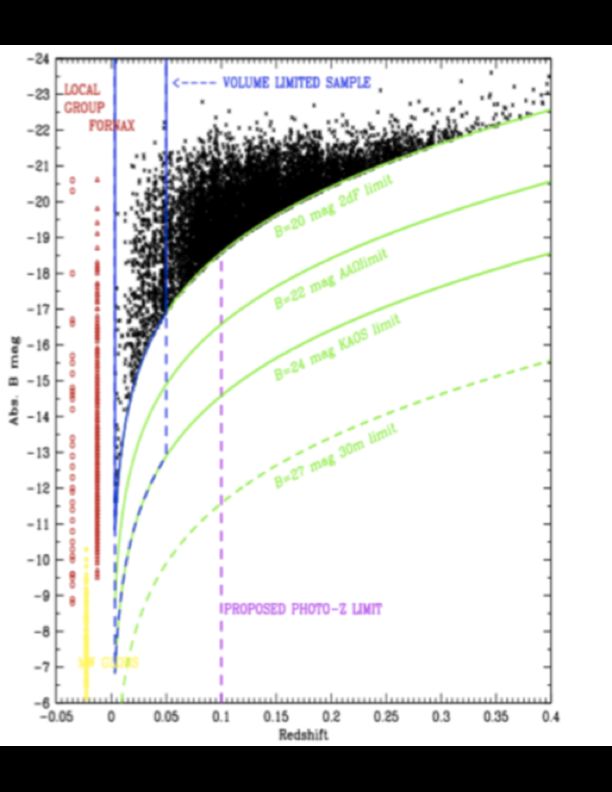}

	\vskip 22pt \caption[The lower end of the luminosity function]{\textbf{The lower end of the luminosity function.}  \emph{The yellow box marks the boundary of the sample limited at $\rm{B} = 20$ mag.  This box barely contains volume for very low luminosity systems.  On the left are seen the distribution of galaxies from the local group, Fornax, and the Galaxy's globular clusters. Image borrowed from Driver~(2004).}}
	\label{diff2quantif}
	\end {center}
\end{figure}

The relative number of galaxies of various Hubble types is usually represented in terms of the luminosity function (LF) as shown in Figure~\ref{globlum}.  The galaxy LF is closely related to the galaxy mass function, one of the most important parameters in galaxy formation models \cite{faintendgal}.  The LF (loosely based on the Press-Schechter formalization for the primordial halo distribution) has for nearly 30 years been a standard tool for quantifying the galaxy population \cite{simbey} and is a useful description of the galaxy content of any particular environment as it is straightforward to measure \cite{faintendgal}.  Recent  studies consistently show that, with some exceptions, low surface brightness (LSB) is synonymous with low luminosity \cite{simbey}.  Small stellar fraction and very low luminosities make DGs the hardest galaxies to detect.  It is in fact generally problematic to quantify the low end of the LF as illustrated by the lack of points contained under the low luminosity area of the curve in Figure \ref{diff2quantif}
.

The LF can be defined as the number density of galaxies per unit luminosity \cite{faintendgal}. A functional form of a general analytic fit to galactic LFs, as proposed by Schechter \cite*{schechter}, is: 
\begin{equation}
\label{schechter1}
\Phi(L)= (\Phi^*/L^*)(L/L^*)^\alpha e^{-L/L^*},
\end{equation}
\noindent or  equivalently, in terms of absolute magnitude:
\begin{equation}
\Phi(\rm{M}) = (0.4 \rm{ln} 10)\Phi^*10^{0.4(\alpha +1)(M^*-M)} \exp[ -10^{0.4(M^*-M)}],
\end{equation}
\noindent where $\alpha$ and $L^*$ (or $\rm{M}^*$) are free parameters used to obtain best possible fits to available  data \cite{co}.  The normalization at the  characteristic luminosity $L^*$ is $\Phi^*$; it is used as such a free parameter in some descriptions of the LF.  Subdividing the morphological galaxy types finely, shows that different classes of galaxy have very different LFs (BM98).  The LF consistently provides a good formal fit to the observed luminosity distribution.  This consistency between the luminosity distribution and the LF appears to hold regardless of environment.  A value of $\alpha=-1$ implies equal numbers of galaxies in the magnitude intervals, a more negative (or steep) value implies that dwarf systems are more numerous \cite{simbey}.

To explain the shape of the LF in a cold DM Universe, it is necessary to include at least a feedback mechanism (beyond the heating resulting from the photo-ionization of the pregalactic gas) to  flatten the faint end of the LF and to suppress cooling at the centers of the massive haloes of groups and clusters. Despite various mechanisms having been attempted, however, it is not a simple matter to replicate the shape; for example, to reproduce a sharp cutoff as observed at the bright end of the LF; Benson et al. \cite*{benshapl1a12}.

The question concerning the LF at the faint end is an old problem noted by Zwicky in a private communication to Parolin et al. (2003):  ``Where does it [the faint end of the LF] end and what kind of objects populate it?"  This problem has become easier to address with the high resolution imaging and spectra in modern times, and yet ``fainter than $\rm{M_R}=-16$, the LF is only known in a handful of environments"  \cite{faintendgal}.

  Barkhouse et al. \cite*{bcglowz}, analyze a very large sample of cluster galaxies with the same purpose as Parolin et al. (2003) namely to examine the dependency of the faint end slope on cluster morphology as well as to examine the gradient, if any, in the faint end as one moves towards the outskirts of a cluster.  They do not go very deep in magnitude ($\rm{M_R}<-16$), but their results offer positive evidence.  There has been a suggestion that the faint-end slope of the galaxy LF may be steeper than previous studies indicated.  This suggestion arises from the existence of very faint DGs not being included, such as those having very flat cores and then steep fall-offs (therefore being very hard to detect).  Including these dwarfs would increase the faint-end slope beyond $-1.5$.  Measurements  of tidal radii and carbon star velocity dispersions in Local Group dwarf spheroidals indicate that the largest dynamical $\rm{M/L}$ belong to the least luminous galaxies, thus the $\rm{M/L}$ of LSB DGs could be large.  The $\rm{M/L}$ numbers in the local group range from 5\,\textendash\,10 for Sculptor to 50\,\textendash\,100 for Ursa minor \cite{bothunElsb}.

There appears to be no clear sign of a lower limit to the LF of irregular galaxies.  They are, on average, much fainter than the other galaxy types.  The numbers of irregulars grow as one moves to fainter magnitudes.  Their faint nature can be crudely described by a Schechter function (parameters $\rm{M_B} ^*(\rm{Irr})=-15 + 5\log{\rm{h}}$ and $\alpha(\rm{Irr})=-0.3$, where 100\ \rm{h} is the Hubble constant).  A fit to the dE galaxies can be given by a Schechter function (parameters $\rm{M_B} ^*(\rm{dE})=-16 + 5\log{\rm{h}}$ and $\alpha(\rm{dE})=-1.3$). 
Dwarf ellipticals show a LF that is open ended, similarly to the irregular galaxies  (BM98).  Note that in our simulations the Hubble constant is given a value $H_0 = 73.2 \rm{km \ s^{-1} {Mpc}^{-1}}$ consistent with the results of WMAP3.

\begin{figure} [!htp]
	\centering
	\begin {center}
	\hskip-50pt	
		
		\includegraphics[bb=00 00 605 640,clip=true,viewport=22 109 600 687,
                  scale=.71, angle=0]{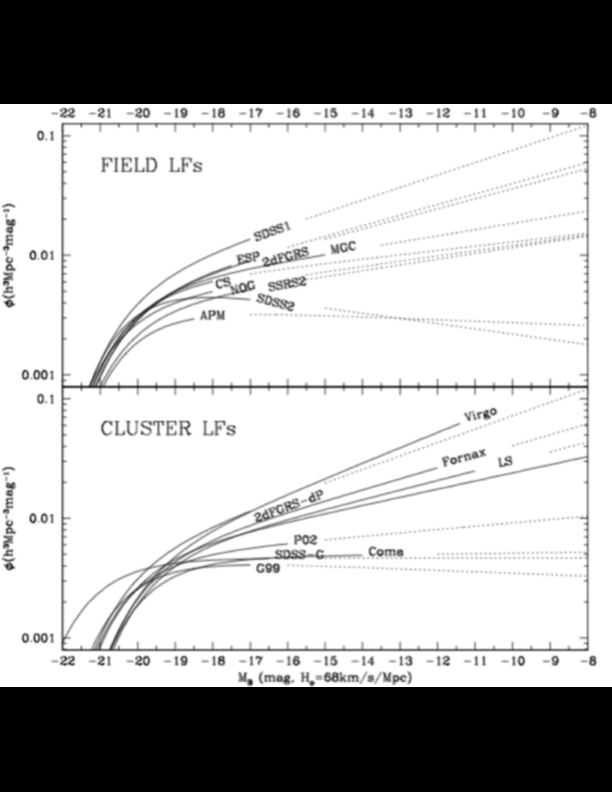}

	\vskip 25pt \caption[Various luminosity functions for different environments]{\textbf{Various luminosity functions for different environments.}  \emph{Data originate from various surveys.  Dotted lines are extrapolations.  Solid lines are data fits.  The cluster LFs are normalized to $\Phi* = 0.0161$.  Shown here is a broad consistence between the field LF and the cluster LF.  Figure borrowed from Driver (2004).}}
	\label{gvsclus}
	\end {center}
\end{figure}

\begin{figure} [!htp]
	\centering
	\begin {center}	
		
				\includegraphics[bb=00 00 605 640,clip=true,viewport=12 39 569 810,
                  scale=.73, angle=0]{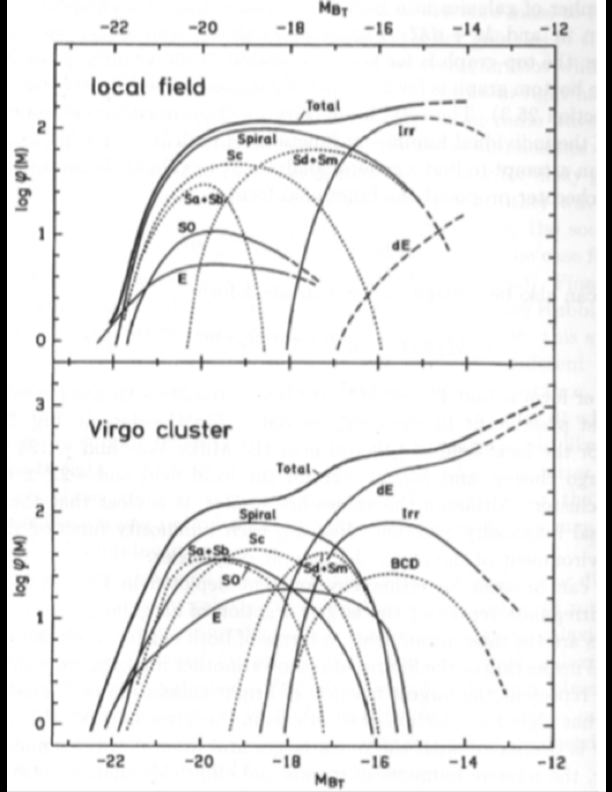}

	\vskip 25pt \caption[Luminosity functions for two samples of galaxies.]{\textbf{Luminosity functions for two samples of galaxies.}  \emph{Image borrowed from Carrol \& Ostlie (1996).  LFs are shown here for samples of galaxies in the vicinity of the Milky Way and the Virgo Cluster (top, bottom, respectively).  The total LF in either environment is the sum of the individual LFs of each Hubble type.}}
	\label{lumfsnearby}
	\end {center}
\end{figure}

\begin{figure} [!htp]
	\centering
	\begin {center}
	\hskip-50pt	
		
		\includegraphics[bb=00 00 605 640,clip=true,viewport=12 103 590 690,
                  scale=.71, angle=0]{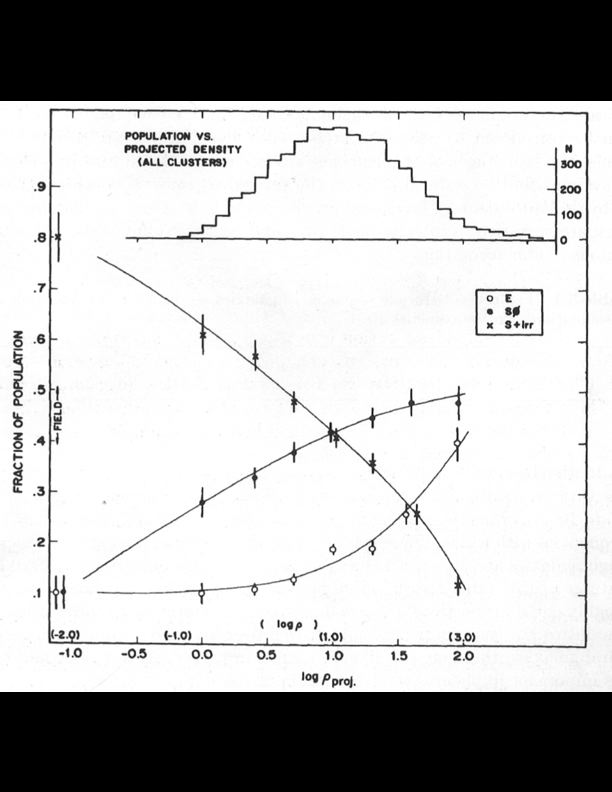}

	\vskip 19pt \caption[Galaxy types in different environments]{\textbf{Galaxy type frequency in different environments.}  \emph{Fraction of morphological galaxy types in different environments (bottom).  Local galaxy number density is shown as projected surface density (top).  This figure was borrowed from Longair~\cite*{gf}}.}
	\label{envirtypg}
	\end {center}
\end{figure}

Although the field and cluster LFs are broadly consistent (Fig.~\ref{gvsclus}), upon comparing the respective LFs for the local field of galaxies near the Milky Way and the Virgo cluster, it is clear that there is not a universal LF.  Rather, each LF depends on the environment of the particular sample of galaxies (Fig.~\ref{lumfsnearby}).  Similarly observations suggest that the galaxy type depends on the environment (Fig.~\ref{envirtypg}).  Astronomers may in the future have the alternative of using information from bivariate distributions as highlighted by Driver (2004), who emphasizes that the global galaxy LF (red line in Fig.~\ref{globlum}) condenses the available information of galaxies (images in the Figure) into three crucial numbers: 1) the characteristic luminosity $\rm{M^*}$; 2) the absolute normalisation $\Phi^*$; and 3) the faint-end slope $\alpha$.  Driver (2004) highlights in addition, that although the Schechter parameterization is usually a remarkably good fit, it may be that too much important information is lost, for instance the sizes and bulge-to-total parameters.

In their long term research, Parolin et al. (2003) have measured a set of clusters in various colors in order to better understand the shape of the LF in clusters and to pinpoint the slope of the faint end.  Their analysis and data interpretation use different algorithms to reduce or eliminate systematic error, robustly corroborating a bimodal nature to the cluster LF.  They emphasize, however, that in their view, it is  the estimate of the total luminosity in different optical color and its correlation with the X-ray luminosity that will give the opportunity, compared with the X-ray luminosity, to better pinpoint some of the characteristics of the cluster and IC medium evolution.

While there is no universal LF, the luminosity function of cluster galaxies is nearly the same from cluster to cluster (Fig.~\ref{gvsclus}), generally following the form proposed by Schechter (1976).  Observing the high luminosity tip of that distribution allows one to normalize the overall galaxy LF for the cluster, yielding estimates for both the cluster's total optical luminosity and its mass.  The number of galaxies in luminosity range $dL$, about $L$, is proportional to $L^{\alpha} e^{-L/L^*} $, with $\alpha\approx-1$ \cite{voitev}.  Carlberg et al. \cite*{dyneqclustgal} showed that the radial mass density $\rho (r)$ and the radial number density $\nu (r)$ of galaxies are roughly proportional to each other; a finding that is used in this project to assign a realistic galaxy distribution for galaxies in the isolated cluster.

The Schechter LF is used variously in this research project to guide the results.  The LF is used, for instance, in combination with a Monte-Carlo rejection function to generate various initial conditions (i.e. the mass of the cluster halo and the mass of the DGs inside the halo).  The first part of the simulation uses galaxy particles with distributions of masses and luminosities chosen to follow a Schechter luminosity distribution and an accepted $\rm{M/L}$.  The initial position of the galaxies (radial coordinate) is then paired to a given galaxy avoiding overlap.  Several series include a central giant cD galaxy.  In the second half of the project presented here, when the simulation has proceeded to an advanced redshift, the mass distribution is plotted and the corresponding LF is compared with observed LFs; this comparison is then used to ensure the obtained results are realistic (Barai et al. 2009). 

\clearpage

\section{Galaxy clusters and the luminosity function }

Clusters of galaxies are a rich information source about the underlying cosmological model; they make possible a number of critical tests.  Because structure grows hierarchically, the growth and development of clusters directly traces the process of structure formation in the Universe.  Clusters are dynamical systems not yet in equilibrium.   Cluster galaxy velocity dispersion and the virial theorem thus cannot yield an exact mass measurement.  Instead, other data (i.e. spatial distribution of galaxy velocities) must be used to accurately measure the masses of nearby clusters \cite{voitev}.

For distant clusters the data necessary to supplement the mass measurements are more difficult to obtain.  To fill this gap, simulations of cluster formation can be used for calibrating the approximate virial relationship between velocity dispersion and cluster mass.  Yet another of Zwicky's \cite*{zwickgrav} important contributions is his idea that cluster masses could be measured through gravitational lensing of background galaxies.  This is one of the primary contemporary methods for measuring cluster mass.  The masses of clusters now are measured at roughly 10$^{15}$ times that of the Sun's.  This tremendous mass notwithstanding, all of the stars in all of a cluster's galaxies represent only a small fraction of a cluster's overall mass.  The majority of the mass, in clusters as throughout the Universe, is an unobserved component, DM. (It is the DM's gravity that caused the baryons to gravitate together into the DM's deep potential wells, there to become galaxies.)  Clusters contain substantially more mass in the form of hot gas than in stars.  This hot gas is observable with X-ray and microwave instruments. A cluster's total mass turns out to be about seven times the combined baryonic mass in stars and hot gas \cite{voitev}.

Models for the X-ray surface brightness of clusters are obtained using the hydrostatic equilibrium equation (variations include assumptions, i.e. spherical symmetry, and isothermality of the gas).  Although clusters are dynamical systems that have not  yet completely equilibrated, they appear to be in approximate hydrostatic equilibrium.  These models are viable because of the relationship between the gas temperature (obtained from the X-ray spectrum) and the depth of the cluster's potential well; the deep potential wells compress the gas, causing it to emit in the X-ray band \cite{voitev}.

As per Voit (2005), under the assumption of spherical symmetry there is, as a function of the radius from the density center, a relation between the gas density $\rho_{g}$ and the temperature $\rm{T}$:

\begin{equation}
\frac{d\ln\rho_g}{d\ln \textit {r}} + \frac{d\ln \rm{T}}{d\ln\textit {r}}=-2\frac{\rm{T}_\phi(\textit{r})}{\rm{T}}.
\end{equation}

\noindent The characteristic temperature $\rm{T}_\phi$ of a singular isothermal sphere with the same value of $\emph{M}({r})/{r}$ is given by:
\begin{equation}
\rm{k_B}\rm{T}_\phi(\emph{r})=G\emph{{M}(r)}\frac{\mu {\emph m_p}}{2\emph r},
\end{equation}

\noindent where $\rm k_B$ is the Boltzmann constant, $\mu$ is the mean molecular weight of the IC gas, and $m_p$ is the proton mass.  The well-known $\rm{\beta}$ models add the assumption  that the gas is isothermal \cite{xhotplasmc,voitev}.  
Assuming an isothermal velocity distribution and a constant velocity dispersion $\sigma_{1D}$, gives:
\begin{equation}
\frac{d\ln\rho_g}{d\textit {r}} = -\frac{\mu m_p}{\rm kT}  \frac{d\phi}{d\textit {r}}  = \beta \frac{d\ln \rho}{d\textit {r}},
\end{equation}
where  $\rho_g$ is the gas density, and $\rho$ and $\rm{T}$ are generalized to the particles responsible for the mass distribution.  Further, as detailed in Voit \cite* {voitev}:
\begin{equation}
	\beta \equiv \mu m_p \sigma^2_{1D}/ \rm{kT}.
\end{equation}
The observed surface brightness profiles of clusters is described quite well by the $\beta$ models in a given radial range, but give the best fits for rich clusters and a possible trend toward lower $\beta$ values in poorer clusters.  Most of the observed X-rays must come from a relatively small proportion of the IC medium (as the X-ray luminosity integrated over radius converges for $\beta>0.5$; Voit 2005).  The $\beta$ models can underestimate the central surface brightness within the core radius $r_c$ and overestimate the brightness at $s \gg~r_c$
(Vikhlinin et al. 1999).
This is in part because of the assumption of isothermality of the IC medium (an idealization), and in part because real cluster potentials differ from the King model \cite{voitev}.

There is a generic form for representing density profiles shallower than isothermal at small radii and steeper than isothermal
at large radii, as corresponding to DM haloes; this is:
\begin{equation}
{\rho_M(r)\propto r^{-p}(r + r_s)^{p-q}}_,
\end{equation}

 \noindent where $p$ and $q$ describe inner and outer power-law slopes, correspondingly,  and the radius $r_s$ marks the steepening of the profile \cite{voitev}.
 \noindent Some models for the X-ray surface brightness include the $\beta$ model, the King model ($\beta=1$) \cite{1998A&A...338..874M}, and the Navarro, Frenk \& White model \cite{NFW97}, hereafter NFW, as well as the Hernquist model \cite{cdmprofiles}.

The $\beta$ model is given by:
\begin{equation}
\label{rho-beta}
{\rho_{\rm gas}\left(r\right)
=\rho_0^{\phantom2}\left[1+\left(r/r_c\right)^2 \right]^{-3\beta/2}}_,
\end{equation} 
\\
where $\rho_0$ is the central density.  
The values of $\rho_0^{\phantom2}$, $r_c$ and $\beta$ are taken from Piffaretti \& Kaastra (2006a),
who give the gas density parameters for 16 nearby clusters.
Scaling the gas density with the universal ratio of matter (dark + baryonic) to baryons gives the halo density:
$\rho_{\rm halo}^{\phantom2} 
=\rho_{\rm DM}^{\phantom2}+\rho_{\rm gas}^{\phantom2}= 
\rho_{\rm gas}^{\phantom2}\Omega_{M}/\Omega_{\rm b}$, where $\Omega_{ M}$ and
$\Omega_{\rm b}$ are the present matter (baryons + dark matter) density parameter 
and baryon density parameter, respectively. 
 The foregoing assumes that the cluster baryon mass fraction follows the cosmic value of $\Omega_{\rm b}/\Omega_{M}$; this is expected to be generally true, although precise estimations of cluster baryon content have shown deviations from the universal value \cite{white93,ettori00,gonzalesz07}.

The much used NFW model is a special case ($\xi=1$) of the functional form as in  Biviano \& Girardi \cite*{biviano03}:
\begin{equation}
\label{rho-anal}
\rho^{\phantom2} \left(r\right)=\frac{\rho_0} {\left(r/a\right)^\xi
\left(1+r/a\right)^{3-\xi}}\,.
\end{equation}
The NFW model is obtained when the distribution of gas and DM in the background halo follow analytical models of the DM density having a functional form as:
\begin{equation}
\label{rho-anal}
\rho_{\rm DM}^{\phantom2} \left(r\right)=\frac{\rho_s} {\left(r/r_s\right)
\left(1+r/r_s\right)^2} \ ,
\end{equation}
\noindent where $\rho_s$ is a scaling density, and $r_s$ is a scale length  \cite{NFW97}.
The NFW profile is often parametrized in terms of a concentration
parameter $c$. The parameters $\rho_s$ and $r_s$ are then given by
\begin{eqnarray}
\rho_s&=&{200c^3\rho_{\rm crit}(z)\over3\left[
{\ln\left(1+c\right)-c/\left(1+c\right)}\right]}
={25H^2(z)c^3\over\pi {\rm{G}} \left[
{\ln\left(1+c\right)-c/\left(1+c\right)}\right]} \ ,
\\
r_s&=&{r_{200}^{\phantom2}\over c}\,,
\end{eqnarray}
\noindent where
$\rho_{\rm crit}(z)=3H^2(z)/8\pi \rm G$
 is the critical density at formation redshift $z$, and
$r_{200}$, the virial radius, is the radius of a sphere whose 
mean density is $200 \rho_{\rm crit}$ (200 times the critical density 
of the Universe at the epoch of formation). After scaling, the halo
density profile is
$\rho_{\rm halo}^{\phantom2} 
= \rho_{\rm DM}^{\phantom2}\Omega_{M}/(\Omega_{M}-\Omega_{\rm b})$.

The Hernquist model differs only in the exponent of the numerator, as below:
\begin{equation}
\label{1rho-anal}  
\rho_{\rm DM}^{\phantom2} \left(r\right)=\frac{\rho_s} {\left(r/r_s\right) 
\left(1+r/r_s\right)^3}\,.
\end{equation}

\noindent The mass density profile can also be approximated by:
\begin{equation}
\label{2rho-anal}
\rho \left(r\right)=\frac{\rho_0} {\left(r/r_s\right)^{1.4} 
\left(1+r/r_s\right)^{1.6}}\,,
\end{equation} 
with $\rho_0\approx5900\rho_c$ and $r_s=370$ kpc.
While the outer-region slope ($\alpha\approx$  3) is steeper then as given by Moore et al. (1998), both of these profiles agree as to the inner profiles\\ ($\alpha=1.4$; Lewis et al. 2000).  More recent high resolution simulations all show $\sim  r^{-1.5}$ for the inner profile, turning over to $r^{-3}$ at large radii  \cite{effectgasdynetcCF,moorbook}. The above equation \ref{2rho-anal} matches the density profile of NFW at large radii but has a steeper central cusp.

The simulations of this project use the above profiles ($\beta$ and the NFW profiles) to model the isolated cluster and subsequently the clusters in the simulated cosmological volume.  In the first case ($\beta$ model) it is assumed that the DM in the background halo follows a density distribution similar to observations of the IC gas.  The halo density is obtained by scaling the gas density with the universal ratio of matter (dark + baryonic) to baryons.  The density profile is then integrated to obtain the background halo mass.  In a second case the halo mass is obtained by integrating over the NFW profile, with the assumption that the distribution of gas and DM in the background halo both follow the NFW profile.  In a future project it might be possible to use a more detailed profile, perhaps bimodal, as suggested by the results of several researchers as below.

Parolin et al. (2003) find consistency between their results and other works in the literature confirming the bimodal nature LF of cluster galaxies.  Driver~et~al.~\cite*{driv94}, studying Abell 963 ($z \approx$ 0.2), in their analysis on a sample that reaches $\rm M_R = -16.5$, similarly find a LF that could be fit by a composition of two Schechter functions, or by a Schechter function plus a power law with slope $-1$ and $-1.8$ respectively (in reasonable agreement with other findings).  Supportive results are also found by Molinari~et~al.~\cite*{1998A&A...338..874M}; and by Barkhouse~et~al.~(2002), who analyze a very large sample of cluster galaxies as well as the LF of the first cluster of the sample used by Parolin~et~al.~(2003); namely their results corroborate a bimodal LF and a faint ~end slope, $\alpha= -$1.65 \cite{parolinlfclust}.

It is possible that the LF condenses too much into three simple numbers, thus losing valuable information.  Driver \cite*{simbey} stresses that the community should use the legacy datasets to supercede the LF with for example two multivariate distributions: (bivariate) the luminosity-surface brightness plane (LSP), and the color-luminosity plane. Galaxy bulges and galaxy disks form marginally overlapping but distinct distributions in both planes.  This indicates two key formation/evolutionary processes: merger and accretion.

The luminosity-surface brightness relation and the luminosity size distribution denote two LSPs.  They are each readily transformed into the other, luminosity, size and surface brightness being related by:
\begin{equation}
\mu^e_{\rm HLR} = \rm{M} + 2.5 \log_{10} [2\pi R^2_{\rm HLR}] + 36.57, 
\end{equation}
\noindent where $\mu_{\rm e}$ is the effective surface brightness, $\rm{M}$ the absolute magnitude, and $\rm R_{\rm HLR}$ the semimajor axis half-light radius in kpc \cite{simbey}.

The LSP offer a strong argument for their utility: both planes show that disks and bulges form distinct but overlapping distributions thus indicating secular evolution of these components (two mechanisms and two timescales).  The LF is a useful global measurement of cluster properties, but this finding presents a strong motivation to begin measuring distinct components independently.  
The LSP appears to provide a direct meeting ground to the numerical simulations.   This is a most important result as it is the cross-talk between simulations and observations that will ultimately bring about the real insights into the processes of galaxy evolution and formation \cite{simbey}.  Driver (2004) further raises three questions as worthy of attention as the community engages, as he suggests, in quantifying the evolution of the LSP distributions across the entire path length of the Universe:
``1) Which wavelength is optimal for structural studies of galaxies? 
2) How might we push back the boundaries into the dwarf regime? and 
3) Can we connect structural measurements to the properties of the DM halo?"

Tests concerning the internal structure of the DM halo are beyond the scope of this project, but in other simulations aiming at explaining, for example the origin and nature of the cusp problem, the bimodality of cluster haloes, and other features of the DM structure, a bivariate distribution such as the LSP might be a useful modeling tool in a parallel way as the LF is used herein.

\chapter{The Isolated Cluster: Method}

With a simple FORTRAN algorithm we simulated (243 simulations) the evolution of an isolated cluster of galaxies in order to examine the relationships between the different methods of DG destruction. 

Careful selection of initial conditions, parameters, and simplifications was essential for the successful implementation of the simulation of the isolated cluster.  These simplifications and assumptions were deemed to be acceptable as they do not significantly affect the average outcome of the simulations.

The assumptions regarding the cluster simulations are as follows:

\noindent$\bullet$Isolated

\noindent$\bullet$Completely formed (comprising $N$ galaxies, represented each by 1 particle, orbiting inside a background halo of uncollapsed DM and gas)

\noindent$\bullet$Experiencing neither mergers with other clusters nor accretion of intracluster matter

\noindent$\bullet$Comprised of a stationary (i.e. non-responsive to forces) spherically symmetric DM halo, the density profile of which does not evolve with time 

\noindent$\bullet$Galaxies will incur one of six possible fates:

   (1) galaxy-galaxy merger
     
   (2) destruction by larger galaxy tidal field, with fragments accreting to larger galaxy
     
   (3) destruction by tides of larger galaxy, with dispersion into the IC medium

   (4) destruction by background halo's tidal field

   (5) galaxy survives to present
     
   (6) ejection from cluster
	
This part of the project simulates isolated, virialized clusters thus the simulations are not ``cosmological"\kern-2pt. This not withstanding, the particular choice of cosmological model enters the picture twice: in the determination of the radii of galaxies, and in the calculation of the elapsed time between the initial and final redshifts of the simulation.

For the cosmological model the assumptions made are as follows:

\noindent$\bullet \Lambda$CDM model

\noindent$\bullet$Matter density parameter, $\Omega_{M}=0.241$

\noindent$\bullet$Baryon density parameter, $\Omega_{\rm b}=0.0416$ 

\noindent$\bullet$Cosmological constant, $\Omega_\Lambda=0.759$

\noindent$\bullet$Hubble constant, $H_0=73.2$ $\rm{Km} \,\,{\rm s}^{-1}{\rm {Mpc}}^{-1}$ ($h=0.732$) 

\noindent$\bullet$Primordial tilt, $n_s=0.958$, and

\noindent$\bullet$Cosmic microwave background temperature, $\rm{T}_{\rm CMB}=2.725{\rm K}$, consistent with the results of WMAP3.
	
The remainder of this chapter details the author's contribution to the development of the simulations of the isolated cluster.  The following chapter (Chapter 3) details the author's role in the development of a cosmological simulation that extends the results of the isolated cluster to a cosmological volume.
	
\section{Code and tests}

In preparation for the simulations it was necessary to ascertain that the research project had not already been carried out elsewhere.  I carried out extensive searches in parallel with other team members.  These searches were carried out using various search engines such as NASA's Astrophysics Data System (a free tool available on the Internet), arXiv.org, and annualreviews.org, among others.  Articles deemed to be relevant were summarized for subsequent group discussion.  This search was continued beyond the initial stages of the project.
 
While conducting the literature search, the members of the research team examined the particle particle (PP) code to be used.  
The PP code is written in Fortran 77.    Fortran is used in astrophysics simulations because of its speed with float operations.  Additionally, legacy code is often written in this language as translating to something more modern, such as Fortran 90 would involve much unnecessary work.  The standard PP code evolves a system of $N$ gravitationally interacting particles using a second-order Runge-Kutta algorithm. It was to be used for the simulation of the isolated cluster after modifications by the team to include interaction with the background halo, and the additions of modules to deal with the subgrid physics.  In the resulting, modified, algorithm the number of particles $N$ can vary, as the particles (galaxies) merge, are destroyed by tides, or escape the cluster \cite{baraidgevol}.

 \begin{figure} [!ht]
	\centering
	\begin {center}	
		
		\includegraphics[bb=00 00 605 640,clip=true,viewport=12 159 600 630,
                  scale=.75, angle=0]{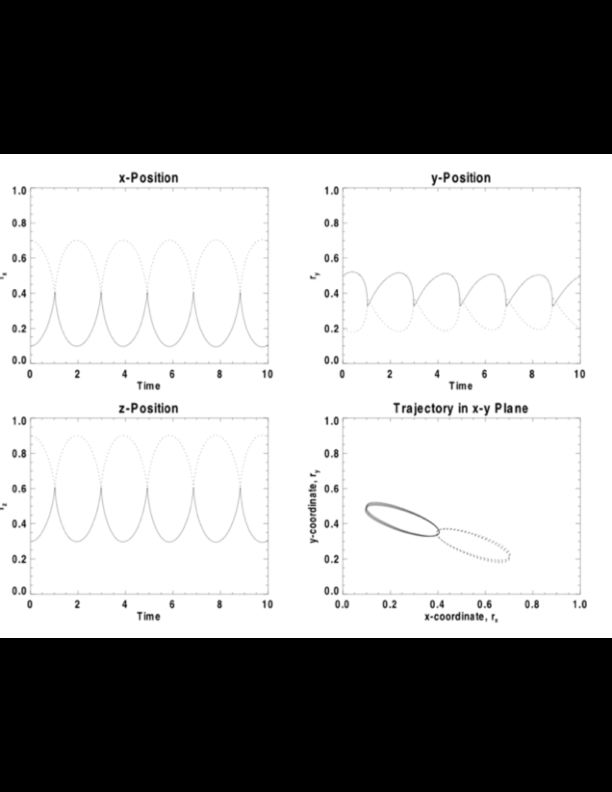}

		\vskip-20pt \caption[Two particle test encounters precession: X-Y plane]{\textbf{Two particle test encounters precession: X-Y plane.}  \emph{Motion of the two test particles is shown on the right bottom panel for x-coordinate versus y-coordinate positions.  Other panels: x-, y-, and z- position versus time.  Precession is visible in the box showing the trajectory in the X-Y plane. }}
	\label{1twopartest}
	\end {center}
\end{figure}

The system we model in this project is an isolated cluster comprising $N$ galaxies of mass $m_i$, radius $s_i$, and internal energy $U_i$ ($i=1,\ldots,N$), orbiting inside a background halo of uncollapsed DM and gas.  For many simulation runs cluster properties similar to well observed clusters (Virgo, Perseus) were used.  Galaxies in the code are represented using one particle per galaxy.  By comparison with contemporary simulations (i.e. $10^6$ particles), this is a very small number.  The use of Ôsubgrid physics,Õ embodied in various subroutines added subsequently, made it possible to use such a relatively small number of particles.  In turn this permitted the use of a direct method (such as the PP code) instead of a more complicated mesh algorithm, for example.

\begin{figure} [!ht]
	\centering
	\begin {center}	
	\hskip-1truecm	
	
		\includegraphics[bb=00 00 605 640,clip=true,viewport=12 179 600 610,
                  scale=.77, angle=0]{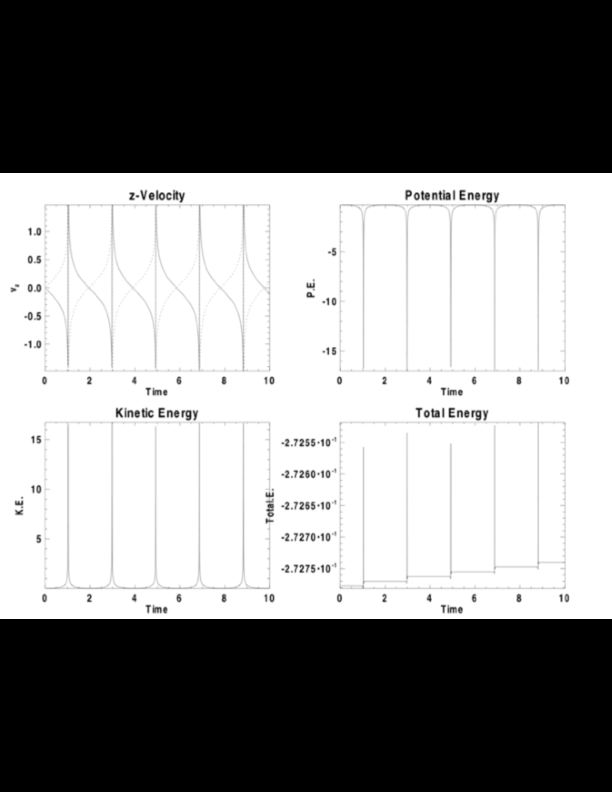}

	\caption[Motion of the two particles in the X-Z plane.]{\textbf{Motion of the two particles in the X-Z plane.}  \emph{Top left panel: z-velocity versus time; top right panel: potential energy versus time; bottom left panel: kinetic energy versus time; bottom right panel: total energy versus time.  The velocities and energies varied in time as  expected, with the kinetic energy reaching a peak as the potential energy is at a minimum.}}
	\label{twoparxz}
	\end {center}
\end{figure}

Before carrying out the actual simulations, it was decided that tests were necessary to ensure the simulation would begin on strong footing.  These tests were carried out for two particles in orbit and for a collapsing uniform sphere.

Plugging two particles into the PP code yielded a test  simulating the two particles orbiting each other around a common center of mass.  \\ \\The energies and velocities of the particles were as expected, and the particles orbited smoothly around a common center but the orbits precessed when the minimum distance of approach between the particles was shorter than a given distance, the softening  length $\epsilon$ (Figs. \ref{1twopartest} 
and \ref{twoparxz}).  The softening length in this test was necessary to prevent an infinite force as these point particles approached each other (in this test we had set $\epsilon=0.1$).  This is important in the subsequent calculation of the force each particle exerts on other particles during the simulation.

The results of this test were analyzed jointly.  The consensus for this test was that the softening length was too large and that this led to particle overlap and the ensuing orbital precession visible in the figures.  Reducing the softening length ($\epsilon=0.0005$) also corrected a similar problem in the kinetic energy plotted for the particles.

In the second test particles collapsed (in the free-fall time) under gravity, from an initial spherical distribution centered at the origin (Figs. 2.3\,\textendash\,\ref{energevol}).  Although the particles collapsed smoothly, they bounced towards the end of the collapse.  At a time close to the free-fall time, there was a reversing in the magnitudes of the potential and kinetic energies in the related plots (Fig. \ref{energevol}).  Tuning $\epsilon$ eventually produced an ideal value of  $\epsilon=0.001$.  It was then decided that in the main simulation $\epsilon$ would be of order of the radius of the smallest galaxies.

\begin{figure}[ht]
     \centering
     \subfigure{
          \label{colinit}
          \includegraphics[viewport=0 0 545 365,clip=true,width=.45\textwidth]{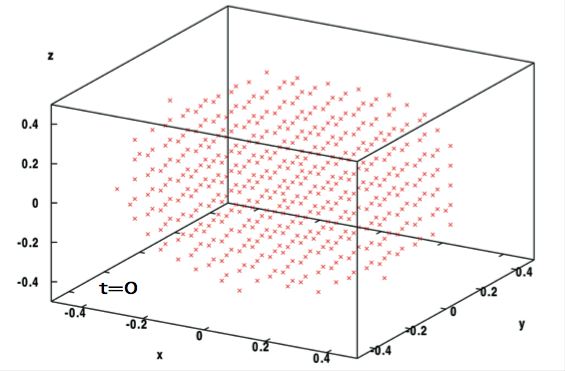}}
     \hspace{.3in}
     \subfigure{
          \label{2colmid}
          

          \includegraphics[viewport=0 0 535 365,clip=true,width=.45\textwidth]{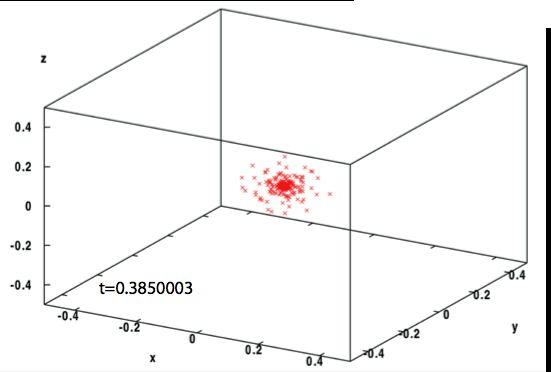}}\\
     \vspace{.3in}
     \subfigure{
           \label{3colmid}
           \includegraphics[viewport=9 0 555 405,clip=true,width=.45\textwidth]
                {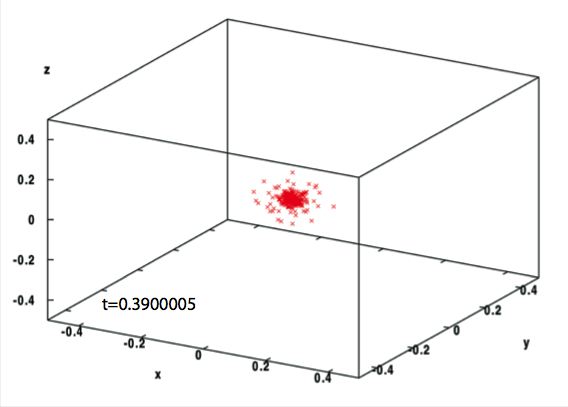}}
                     \subfigure{
           \label{4colmid}
             \hskip 19pt
             \includegraphics[viewport=0 0 555 415,clip=true,width=.45\textwidth]{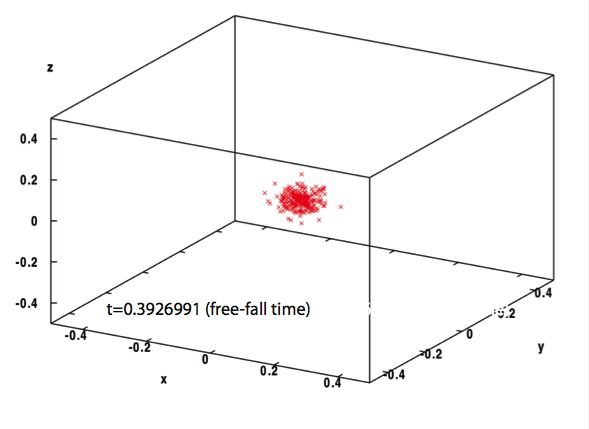}}
     \vskip 27pt \caption[Spherical collapse test; initial positions]{\textbf{Spherical collapse test; initial positions.}  \emph{This test was done for 503 particles cubed.  The total time is the free-fall time.  At $\rm{t}=0.385$ the first particles reach the center.  At $\rm{t}=0.3926$ the free-fall time is reached.  Some particles do not reach the center; some particles bounce back.}}
     \label{fig:2858multifig}
\end{figure}

 \begin{figure} [!ht]
	\centering
	\begin {center}
	
	\includegraphics[bb=00 00 605 640,clip=true,viewport=02 211 600 585,
                  scale=.74, angle=0]{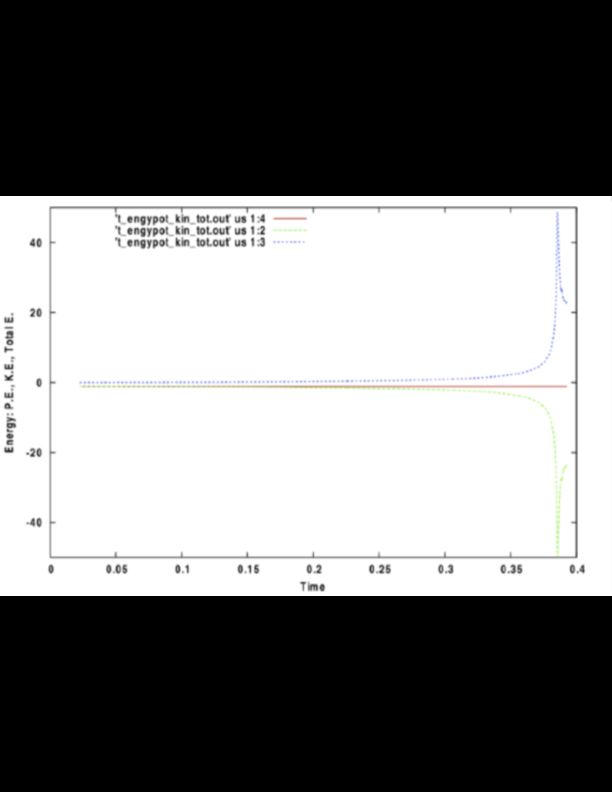}

	\caption[Spherical collapse test; energy is conserved in system]{\textbf{Spherical collapse test; energy is conserved in the system.}  \emph{This plot shows the evolution of energy of the system of particles under spherical collapse.  For this test the softening length $\epsilon=0.001$.  The kinetic energy (short-dash line) plus the potential energy (long-dash line) add up to zero with energy conservation (middle, solid line).}}
	\label{energevol}
	\end {center}
\end{figure}

\clearpage

\section{Parameter searches}

Several parameters were required to describe the cluster, these are related to: halo and galaxies, softening length, galaxy interactions, and to general simulation components.  It was decided that cooling flows, accretion to cluster, and galaxy formation would be ignored (subsequently a treatment of cluster accretion was implemented).  Important parameters to find in the literature were:

\begin{itemize}
\item Mass or density cluster profile  (from either X-ray luminosity or radial velocities)	
\item Number of galaxies to be used
\item Galaxy mass spectrum $N(m)$ (post-search, Schechter LF is used)
\item Relation mass-size for galaxies (observational and theoretical)
\item Initial positions and velocities of galaxies (caveat: massive galaxies near center)
\item Cluster formation epoch
\end{itemize}

\subsection{Density profile}

There was an open question as to which density profile would be used in the simulations, but as it was easy to model several by slightly modifying the code, it was decided that the $\beta$ and the NFW models for the density profile of the cluster would be used.  I had also found that the  $\beta$ model is widely thought to overestimate the mass, especially at large radii.  Modeling various density profiles would allow comparison of the various results.

\subsection{Cluster mass}

        The force exerted by the halo is proportional to $M(r)/(r^2+\epsilon^2)^{3/2}$, where $M(r)$ is the mass inside radius $s$, and $\epsilon$ is the softening length of the force.  The original plan had been to tabulate this quantity as a function of the radius.  It became apparent that each density profile would need a separate interpolation table for each value of  the softening length
 .  To reduce the number of interpolation tables it was decided to  tabulate the quantity $M(r)$ only, with the division by $(r^2+\epsilon^2)^{3/2}$ being performed during the simulation.

The code to make the tables {\tt{maketable.f}} is presented in Appendix \ref{appinterp}.  This code was written by Dr. Martel and modified by myself.  The code facilitates calculation of i.e. the $\beta$, NFW, King, and Hernquist mass profiles.  In order to make the tables it was necessary to integrate over the function determining the desired profile.  In {\tt{maketable.f}}, $Smax$ represents the distance from the center, out to which it is desired to carry out the integration; whereas $ds$ is the incremental step the integration procedure will use for the iterations.  The number of entries to the table corresponds to $npt$; this in turn gives the level of definition of the produced table. 

In {\tt{maketable.f}}, I modified sections of the code that included the algorithm for integration.  In order to decide on the boundaries (i.e. a distance for $Smax$, or a limit for $npt$) I conducted several trial runs until the team had decided on the optimal values.  It was an interesting point in the project to find and define a physical correspondence (parameters $Smax$, $ds$, $npt$) between the cosmological object and numerical methods for integral calculus.  For instance, to correctly produce the interpolation tables, I had to define $ds=Smax/npt$.

\subsection{Initial conditions}

\begin{figure} [!htp]
\begin{center}
\includegraphics[width=6in]{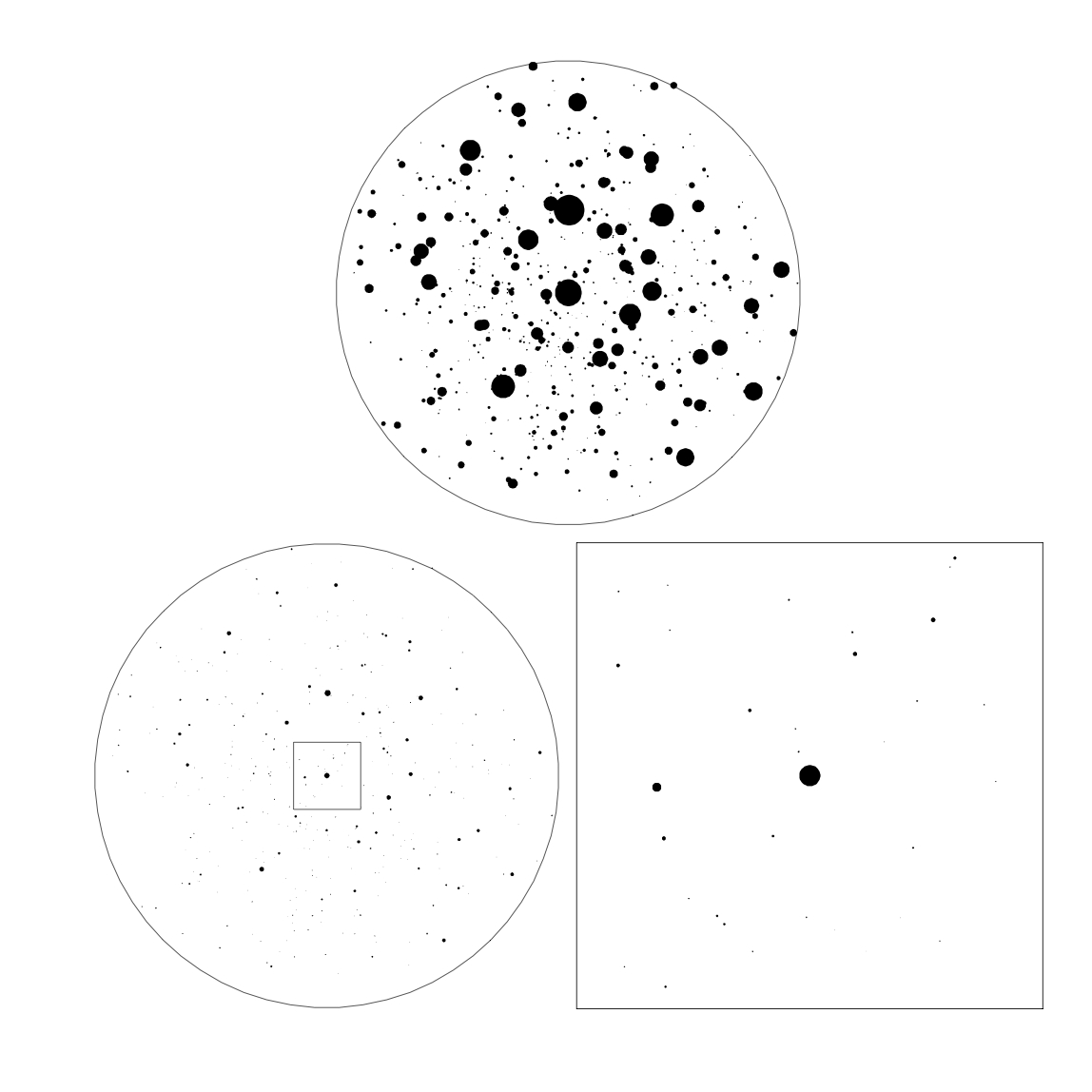}
\caption[Initial conditions for run A12]{\textbf{Initial conditions for run A12.} Top panel:
initial conditions at $z=1$. \emph{The solid circles indicate the
virial radii of galaxies. The large circle is the maximum distance 
$r=3R_0=2.08\,\rm Mpc$
from the cluster center. Lower left panel: same as top panel, 
with symbols rescaled to optical diameter of real galaxies. 
Bottom right panel: enlargement of the central $(0.6\times0.6)\,{\rm\,Mpc}^2$
box on lower-left panel.}}
\label{init_A12}
\end{center}
\end{figure}

It was necessary to choose the initial mass $m$, radius $r$, position $s$, and velocity $\bf v$ of each galaxy in order to determine the initial conditions for the simulations.

The galaxies were assumed to follow a Schechter distribution, such as  in equation~\ref{schechter1}.  From this distribution, a Monte Carlo rejection method was used to select the luminosities of the galaxies.  The masses were obtained using these luminosities by means of an assumed $\rm{M/L}$.

To obtain the slope of the Schechter function of the galaxies in the simulations presented herein, it was necessary to plot the relevant luminosities within each run.  I ran a code that extracted the relevant values and I tabulated the results in a table to be presented to the rest of the team.  The relevant values were then extracted so as to fit the slope of the Schechter function as further detailed in the related article (Barai et al. 2009; see Appendix \ref{appart}).

Once the initial conditions were decided, see Figure \ref{init_A12} for an example; and once the various modules of the PP code were in order, the last test run became the first successful simulation run A.

The locations of galaxies inside the cluster were selected with the assumption that their distribution is statistically isotropic and using observations and conclusions from the literature.  To prevent overlap, however, and as a certain structure is expected in the cluster (i.e. the most massive galaxies probably reside near the cluster center), the most massive galaxy was positioned at the center of the cluster.  The next seven most massive galaxies were placed between radii $R_0$ and $2R_0$ (where $R_0$ is 3 times the radius of the most massive galaxy).  The next nineteen most massive galaxies were placed between radii $2R_0$ and $3R_0$.  The remaining galaxies were placed randomly between radii 0 and $3R_0$.  The radius $s$ of each galaxy is taken to be equal to the virial radius $r_{200}$.  The initial velocity of each galaxy was obtained by taking the velocity a galaxy would have if it were in perfect circular orbit and randomly varying the magnitude within a given range.  To determine the direction of the velocities of the galaxies, a random angle generation technique was followed, similar to that used in generating the galaxy positions.

\section{Subgrid physics}

The subgrid physics are effected by modules added to the PP code.  These modules deal with the six possible outcomes of the DG interactions in the simulations here presented.  These six outcomes are:

\begin{itemize}
\item  The galaxy merges with another
\item  The galaxy is destroyed by the tidal field of a larger galaxy, but the fragments  accrete to that larger galaxy
\item  The galaxy is destroyed by tides of a larger galaxy, and  the fragments are dispersed in the IC medium
\item  The galaxy is destroyed by the tidal field of the background halo
\item  The galaxy survives to the present (i.e. it is not destroyed by any process)
\item  The galaxy is ejected from the cluster
\end{itemize}

\begin{figure} [!hbp]
\begin{center}
\includegraphics[width=5in]{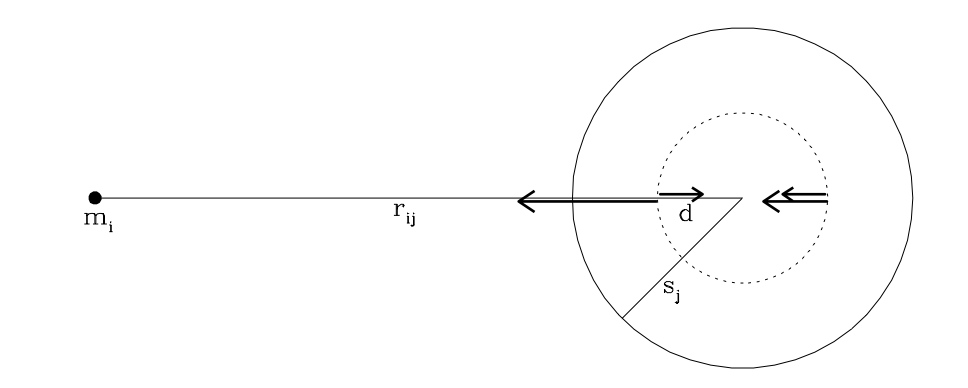}
\vskip20pt \caption[Tides caused by other galaxies]{\textbf{Tides caused by other galaxies.}  \emph{Shown is the effect of tides caused by a galaxy of mass $m_i$ on a galaxy of mass $m_j$ and radius $s_j$. 
The two largest arrows show the gravitational accelerations caused by galaxy $i$; 
the two smallest arrows show the accelerations caused by galaxy $j$.}}
\label{tides_schem}
\end{center}
\end{figure}

Figure~\ref {tides_schem} shows the geometry of the tides due to other galaxies in the simulation.

In the code, the subgrid physics subroutines are named: {\tt{tidesByHalo}}, {\tt{encounter}}, {\tt{tidesByGalaxy}}, {\tt{collision}}, {\tt{OutsideHalo}}, and {\tt{GalxPresent}}.  The module name describes the interaction with which it deals.  I tested these subroutines, written by Dr. Barai, by adding one subroutine at a time, running the code and comparing the ensuing code outputs.  Next the finished module was commented out, the next was added and the process was re-iterated.  When a given subroutine appeared to be operating correctly the next one was added for subsequent testing.  The main steps were adding mergers, tides, and harassment.

The amount of data generated in the run-up tests (and hence in the 243 subsequent runs) would be prohibitive were it not for the tree-structure directories of UNIX and various editing and plotting programs.  I carried out tests for the various modules as they were written, and organized and managed the storage of this data for plotting and comparison.  As the various modules progressed  I ran  the codes to compare the output with previous tests.  In subsequent runs B\,\textendash\,I,  I varied the active subroutine (with or without harassment), cluster properties, and DM profile ($\beta$ or NFW).

\begin{deluxetable}{ccccccccccccc}
\rotate
\tabletypesize{}
\tablecaption{Simulations for series A}
\tablewidth{0pt}
\tablehead{
\colhead{Run} & \colhead{$M_{\rm total}[10^{11}M_\odot]$} 
& \colhead{$N_{\rm total}$} &
\colhead{$N_{\rm merge}$} & \colhead{$N_{\rm tides}^{\rm gal}$} & 
\colhead{$N_{\rm accr}$} & \colhead{$N_{\rm tides}^{\rm halo}$} & 
\colhead{$N_{\rm eject}$} &
\colhead{$f_{\rm surv}^{\phantom1}$} &
\colhead{$f_M^{\phantom1}$} &
\colhead{$f_{\rm ICS}^{\phantom1}$} &
\colhead{$\alpha_{\rm start}^{\phantom1}$} & 
\colhead{$\alpha_{\rm end}^{\phantom1}$} 
}
\startdata
 A1 &  855.0 & 440 & 137 &  54 &  6 & 0 & 1 & 0.550 & 0.225 & 0.235 & $-1.26$ & $-1.23$ \cr
 A2 &  862.0 & 702 & 343 & 182 & 24 & 0 & 1 & 0.217 & 0.490 & 0.522 & $-1.28$ & $-1.12$ \cr
 A3 & 1100.4 & 480 & 126 &  43 &  7 & 0 & 0 & 0.633 & 0.136 & 0.161 & $-1.28$ & $-1.24$ \cr
 A4 &  831.0 & 530 & 165 &  67 & 11 & 0 & 0 & 0.542 & 0.148 & 0.161 & $-1.26$ & $-1.15$ \cr
 A5 &  980.4 & 459 & 175 &  58 &  8 & 0 & 1 & 0.473 & 0.128 & 0.144 & $-1.25$ & $-1.18$ \cr
 A6 &  967.8 & 640 & 230 &  74 &  5 & 0 & 0 & 0.517 & 0.157 & 0.182 & $-1.29$ & $-1.23$ \cr
 A7 &  720.0 & 579 & 263 & 114 &  7 & 0 & 0 & 0.337 & 0.359 & 0.384 & $-1.28$ & $-1.20$ \cr
 A8 &  757.8 & 435 & 200 &  66 & 51 & 0 & 1 & 0.269 & 0.320 & 0.346 & $-1.27$ & $-1.26$ \cr
 A9 &  925.6 & 514 & 223 &  83 & 12 & 0 & 1 & 0.379 & 0.245 & 0.273 & $-1.28$ & $-1.17$ \cr
A10 &  858.9 & 525 & 204 &  54 &  8 & 0 & 0 & 0.493 & 0.257 & 0.260 & $-1.31$ & $-1.24$ \cr
A11 &  880.2 & 547 & 224 &  68 &  9 & 0 & 1 & 0.448 & 0.257 & 0.278 & $-1.33$ & $-1.25$ \cr
A12 &  826.1 & 548 & 255 &  72 & 15 & 0 & 1 & 0.374 & 0.225 & 0.249 & $-1.29$ & $-1.18$ \cr
A13 & 1041.5 & 431 & 166 &  51 &  6 & 0 & 0 & 0.483 & 0.275 & 0.282 & $-1.27$ & $-1.25$ \cr
A14 &  957.3 & 486 & 174 &  52 &  4 & 0 & 0 & 0.527 & 0.261 & 0.265 & $-1.29$ & $-1.22$ \cr
A15 &  860.5 & 520 & 160 &  64 & 15 & 0 & 1 & 0.538 & 0.260 & 0.260 & $-1.26$ & $-1.18$ \cr
A16 &  858.0 & 483 & 255 &  72 &  8 & 0 & 0 & 0.306 & 0.323 & 0.337 & $-1.28$ & $-1.19$ \cr
\enddata
\label{seriesa}
\end{deluxetable}

\begin{deluxetable}{ccccccccccc}
\rotate
\tabletypesize{\scriptsize}
\tablecaption{Series of simulations}
\tablewidth{0pt}
\tablehead{
\colhead{Series} & \colhead{Runs} &
\colhead{$\alpha_{\rm start}^{\phantom1}$} & 
\colhead{Profile} &
\colhead{$\beta$} &
\colhead{$\rho_0^{\phantom1},\rho_s^{\phantom1}\,[\rm g\,cm^{-3}]$} &
\colhead{c} & 
\colhead{$r_c^{\phantom1},r_s^{\phantom1}\,[{\rm kpc}]$} & 
\colhead{cD} &
\colhead{Harassment} & 
\colhead{Cluster Growth}
}
\startdata
 A & 16 & $-1.28$ & $\beta$-Virgo   & 0.33 & $8.14\times10^{-26}$ &$\cdots$ & 
3 & $\times$ & $\times$ & $\times$ \cr
 B & 17 & $-1.28$ & $\beta$-Virgo   & 0.33 & $8.14\times10^{-26}$ &$\cdots$ &  
3 & $\times$ & $\surd$ & $\times$ \cr
 C & 17 & $-1.36$ & $\beta$-Virgo   & 0.33 & $8.14\times10^{-26}$ &$\cdots$ &  
3 & $\times$ & $\surd$ & $\times$ \cr
 D & 16 & $-1.36$ & $\beta$-Virgo   & 0.33 & $8.14\times10^{-26}$ &$\cdots$ &  
3 & $\surd$  & $\surd$ & $\times$ \cr
\noalign{\hrule}
 E & 16 & $-1.36$ & $\beta$-Perseus & 0.53 & $7.27\times10^{-26}$ &$\cdots$ & 
28 & $\times$ & $\surd$ & $\times$ \cr
 F & 16 & $-1.36$ & $\beta$-Perseus & 0.53 & $7.27\times10^{-26}$ &$\cdots$ & 
28 & $\surd$  & $\surd$ & $\times$ \cr
\noalign{\hrule}
 G &  10 & $-1.28$ & NFW & $\cdots$ & $2.35\times10^{-25}$& 5 & 
200 & $\times$ & $\surd$ & $\times$ \cr
 H &  14 & $-1.31$ & NFW & $\cdots$ & $2.35\times10^{-25}$& 5 & 
200 & $\times$ & $\surd$ & $\times$ \cr
I &  10 & $-1.31$ & NFW & $\cdots$ & $2.35\times10^{-25}$& 5 & 
200 & $\surd$  & $\surd$ & $\times$ \cr
\noalign{\hrule} 
 J & 16 & $-1.36$ & $\beta$-Perseus & 0.53 & $7.27\times10^{-26}$ &$\cdots$ & 
28 & $\times$ & $\surd$ & $\surd$ \cr
\enddata
\label{series}
\end{deluxetable}

\section{First results}

 Once the initial conditions had been chosen, and tests were finished in the previous subroutines, the first meaningful simulation run was carried out.  Table~\ref {seriesa} shows the results of this series of simulations. \\This series (series A) constituted 16 simulations, using for the background halo a $\beta$-profile with $\beta = 0.33$, a core radius $\rm{r_c}= 3$ kpc, and a central density $\rho_0 = 8.14 \ee{-26} \rm {g\,cm}^{-3}$, which is appropriate for a cluster like Virgo (Piffaretti \& Kaastra 2006).  This series does not include galaxy harassment, or a cD galaxy.  A total of 10 series were run, comprising parameters as summarized in Table~\ref {seriesa}.    We analyzed the data over several discussions; the resulting explanation is presented in the related article (Barai et al. 2009; see Appendix \ref{appart}).  Some details are excerpted here and summarized in the next chapter.

Table~\ref {seriesa} shows the run number in column 1, and in column 2  the total mass $M_{\rm total}$ in galaxies at the beginning of the run (in units of $10^{11}M_\odot$).  Column 3 shows the number of galaxies $N_{\rm total}$.  Columns $4$ to $8$ show the number of galaxies $N_{\rm merge}$ destroyed by mergers; the number of galaxies $N_{\rm tides}^{\rm gal}$ destroyed by tides caused by a massive galaxy, with the fragments dispersed in the IC medium; the number of galaxies $N_{\rm accr}$ destroyed by tides caused by a massive galaxy, with the fragments being accreted onto that galaxy; the number of galaxies $N_{\rm tides}^{\rm halo}$ destroyed by the tidal field of the background halo; and  the number of galaxies $N_{\rm eject}$ ejected from the cluster, respectively.  Column 9 shows the fraction by numbers of galaxies $f_{\rm surv}$ that survive to the present.  Column 10 lists the values of $f_{M}$, the mass fraction of galaxies contributing to the IC stars.  Column 11 lists the values of the total luminosity of the cluster coming from IC stars, $f_{\rm ICS}$.  Column 12 shows the Schechter luminosity function exponent $\alpha_{\rm start}^{\phantom1}$.  The exponent was obtained by performing a numerical fit to the distribution of galaxy masses. The masses were determined using a Monte Carlo rejection method therefore the exponent can differ slightly from the intended value $\alpha = -1.28$ (equation~\ref{schechter1}).  Averaging over all runs yields $\alpha_{\rm start} = -1.280 \pm 0.020$.
Column 13 lists $\alpha_{\rm end}$, the best fit to the Schechter function for the surviving galaxies in the given simulation.

There is no occurrence of a galaxy being destroyed by tides by the background halo, and the number of galaxies ejected is either 0 or 1. There are large variations in the other numbers from one run to the next, but some trends are apparent. Typically, 50\%\,\textendash\,60\% of the galaxies are destroyed. Run A2 is an extreme case, with 78\% of the galaxies being destroyed. 

The destruction of galaxies by mergers dominated the destruction by tides by more than a factor of 2 except for run A7. If the cases of tidal disruption followed by accretion are treated as being mergers, then mergers dominated tidal disruption even more.  When galaxies were destroyed by tides, the dispersion of fragments into the IC medium always dominated over the accretion of fragments to the massive galaxy, but the ratio varied wildly, from $114:7$ in run A7 to $66:51$ in run A8.

Column 10 of Table \ref {seriesa} lists the values of $f_{M}$,  the mass fraction of galaxies contributing to the IC stars, 
\begin{equation}
f_{M}^{\phantom1}={M_{\rm tides}^{\rm gal}+M_{\rm tides}^{\rm halo}
\over M_{\rm total}-M_{\rm eject}}\,,
\end{equation} 

\noindent where $M$ refers to the mass in galaxies, rather than the number of galaxies. The galactic mass contribution to the IC medium consists of galaxies destroyed by the tides of another more massive galaxy and by the tides of the background halo (though in this series there are no such cases).  The galaxy masses are converted to luminosities using a mass-dependent $\rm{M/L}$ from the cold DM cosmological simulations of \cite{yang03}.

The fraction $f_{\rm ICS}^{\phantom1}$ of the total luminosity of the cluster that comes from IC stars (column 11) is calculated as per:
\begin{equation}
f_{\rm ICS}^{\phantom1}={L_{\rm tides}^{\rm gal}+{L}_{\rm tides}^{\rm halo}
\over {L}_{\rm total}-{L}_{\rm eject}}\,.
\end{equation}

Again, there were large variations; in particular the fraction was very large for run~A2 and very small for run~A5. 
Averaging over all runs we found
\begin{equation}
f_{\rm ICS}^{\phantom1}=0.271\pm0.095\,.
\end{equation} 

\noindent
Even though in most cases about half the number of galaxies were destroyed, the destroyed galaxies tended to be low-mass galaxies; this explained why $f_{\rm ICS}<1-f_{\rm surv}$, for all the runs.

That the galaxies destroyed by mergers and tides or escaping were mostly low-mass galaxies  leads to a flattening of the galaxy mass distribution function.  We computed:

\begin{equation}
\alpha_{\rm end}^{\phantom1}=-1.206\pm0.040\,.
\end{equation}
  
as the best numerical fit to the Schechter LF.

\chapter{The Isolated Cluster: Results}

\section{The series\,\textemdash\,analysis}

The results of the isolated cluster simulations are being published (Barai et al. 2009, see Appendix \ref{appart}).  In this article a detailed account is presented of the simulations.  Following is a description of my contribution to research related to the article, including execution of the numerous simulation runs; also following is a summary of the related results of these simulations.

Table \ref {series}, summarizes the various series of simulations.  The first column in Table \ref{series} shows the series label.  The remainder of each row details the properties of that series.  Column 2 gives the number of runs for each series.  Column 3 gives the value of $\alpha_{\rm{start}}$ for each series.  Column 4 tells which DM profile was used for each series and which cluster that profile resembles.  Column 5 gives the value of $\beta$ used, when applicable.  Column 6 shows the value of the central density.  Column 7 gives, for the NFW profile, the concentration parameter.  Column 8 gives the value of the core radius (the scale radius for the NFW profiles).  Column 9 tells whether or not each run included a cD galaxy.  Column 10 shows whether or not each run included harassment.  Column 11 shows, for each run, whether or not there was cluster growth.  The series thus represent variations of the isolated cluster simulations.  I carried out the simulations for series A, D, E, F, G, H, I, and J.  I also carried out an unpublished $\beta$ profile series with $\alpha_{\rm{start}}=1.34$ and two Hernquist profile series with no cD galaxy and $\alpha_{\rm{start}}=1.28$ and $1.31$, respectively.  It was jointly decided that the related article would present only the NFW and $\beta$ profile related simulations.  

For each simulation run I obtained data that I then tabulated using LaTeX and presented to the rest of the group.  Each series thus contains data similar to Table \ref{seriesa}, which presents the results of the first meaningful simulation run.  The runs in Table \ref{seriesa} simulate a cluster with no massive cD galaxy at its center and with no galaxy harassment.  This series uses the $\beta$-profile for the halo, with $\beta= 0.33$, core radius $r_c = 3 \rm {kpc}$, and a central density $\rho_0 = 8.14 \ee{-26} \rm {g\,cm^{-3}}$.  These values are similar to, for instance, the Virgo cluster \cite{effervheatconst}.

The data obtained after activating the harassment module are given in series B.  From the higher tide related destruction rate it was thought that the algorithms used credibly modeled galactic harassment while maintaining a Schechter distribution of galaxies throughout the simulated evolution of the cluster.

Variations in the parameters were found to have predictable consequences consistent with theory, such as in the increased number of ejections present in series C ensuing when the cD galaxy was particularly massive (which in turn increased the radius in the criterion for initial positions of the galaxies).  Further, using the results of the simulations to this point and observations from the literature we calculated the slope of the obtained luminosity function in order to average and find the best fits.

Incorporating a cD galaxy in the simulation (series D) resulted in an increase in accretions (i.e. by the cD galaxy) and in a lower fraction of IC stars imparted to the IC medium, predictably, as there is less mass available and there are a higher number of accretions. 

Parameters such as those of the Perseus cluster (Piffaretti~\&~Kaastra~2006), using a $\beta$ profile, resulted in tidal destructions by the background halo (series E); this was probably due to the steepness of the cluster density profile in this simulation.  This conjecture is supported by the specifics of the destructions.  A cD galaxy, when present (series F), competes for tidal disruptions with the background halo, although mergers continue to dominate tides.

In series G, H, and I simulated a cluster with NFW profile, while varying the slope of the LF and the presence of a cD galaxy.  In the simulations of NFW clusters, a large number of galaxies were destroyed by the tidal field of the cluster halo. The number of halo-tide destructions is comparable to, or larger than that of the galaxy-tide destructions.  This research team theorizes that this is due to the steepness (within a given radius) of the NFW halo mass profile, resulting in galaxies near the cluster center experiencing a larger tidal field.

Further in series G, with a lower LF slope and no cD galaxy, it can be seen that the increase in halo-tide destructions leads to mergers exceeding the galaxy-tide destructions; combined, however, the tide destructions by galaxy and cluster-halo compare to, or exceed mergers. An associated increase in the mass contributed to the IC stars is also seen.

Series H and I (NFW, including harassment, and having a higher slope LF) are in keeping with the above observations, although there are more accretions in run I, where the cD galaxy is present (similar to series D and F).  When the cD galaxy is present galaxies approaching the cluster center tend to get destroyed by the tidal field of the halo before the cD galaxy can have any effect.  This plays a limiting role in the fraction of matter imparted to the IC medium.

In series J the background cluster-halo is allowed to accrete over time, 33\% from $z=1$ to $z=0$, as per Wechsler et al. \cite*{wechsler02}.  We find that this leads to a higher number of interactions, which in turn decreases the fraction of surviving galaxies.

\section{Cluster profile parameter dependence}

With 60 simulations we explored the parameter dependence of the cluster profiles ($\alpha_{\rm start}~=~-1.36$, with harassment, no cD galaxy).  We explored the range $0.3 < \beta < 1.0$ (at fixed $r_c$ and $\rho_0$), these being typical values as per the literature.  The results of these simulations and the joint analysis are summarized here (for details see Section 5 of the related article of Barai et al. 2009; Appendix \ref{appart}).

\subsection{The $\beta$ profile}

An increase in $\beta$ causes a decline of galaxy interactions, hence an increase in the number of galaxies surviving to $z = 0$.  With mostly the smaller galaxies merging, the number of mergers dominated the galaxy tides, but the merging mass fraction decreased below the tidal destruction mass fraction (at $\beta < 0.6$).  As $\beta$ was increased from $0.3$ to $0.6$, however, it appeared that more massive galaxies were merging.  At $\beta > 0.6$ the galaxy accretion dominates the mass-fraction.

The IC stars luminosity comes dominantly from the DGs destroyed by the tides of more massive galaxies.  Exploring variations of the core radius $r_c$ between 10\,\textendash\,500 kpc (Table 12 in the related article of Barai et al. 2009; see Appendix \ref{appart}) showed that for small $r_c$, mergers outnumber galaxy tides, are comparable to tides at $r_c \approx 350$ kpc, and that tides dominate at large $r_c$.  As $r_c$ rises, a decreasing number of galaxies survive up to $z=0$.  This leads to an increase in the mass contributed to the IC medium.  When $r_c > 50$ kpc, the mass fraction is increasingly dominated by galaxy-tide destruction, with the fragments being dispersed into the IC medium.  The merged mass fraction, however, is significantly smaller than that of tidal destructions as the merging galaxies are mostly low mass.  As $r_c$ increases, galaxy tides increase substantially by number and mass, while mergers remain almost constant.

\subsection{The Navarro, Frenk, and White profile}
The dependence on the parameters governing the NFW model cluster halo density profile was explored with 35 simulations ($\alpha_{\rm start}= -1.31$, with harassment, no cD galaxy). As per the literature, typical values used for the scale radius are $r_s = 100$\,\textendash\,$500$ kpc and for the concentration parameter, $c = 3$\,\textendash\,$6$.

We investigated five different values of the scale radius $r_s$ within 10\,\textendash\,300 kpc, keeping the concentration fixed at $c = 4.5$.  With increase of $r_s$, a smaller number of galaxies survive up to the present, causing greater mass to be transferred to the IC medium.  Near $r_s \geq 100$ kpc, the mass fraction is increasingly dominated by halo tidal destructions.  When $r_s$ is small, the number of mergers is higher than that of tide destructions; but  at $r_s \approx 300$ kpc the numbers of mergers, galaxy-tide, and halo-tide destructions become comparable. The merged mass fraction, however, is comparable to that of galaxy-tide destructions; these are both dominated by halo-tide destruction. 

It seems that as $r_s$ increases, lower-mass galaxies are tidally destroyed by other galaxies; the galaxy-tide destructions decrease in number but increase in mass.

Varying the concentration parameter to $c = 4$ and $6$ with a fixed value of the scale radius $r_s = 100$ kpc, it was found that the mergers continued to outnumber the tides.

\section{Discussion}

\subsection{Mergers and tides} 

In the $\beta$ model, destruction by mergers dominates destruction by tides;  in the NFW model they are of comparable importance. 

The relative importance of mergers over tides is arrived at by calculating the average (over all runs of each simulation) of the fractions below as per the related article (Barai et al. 2009; see Appendix \ref{appart}):

\begin{eqnarray}
f_{\rm destroyed}^{\rm mergers}&=&{N_{\rm merge}+N_{\rm accr}
\over N_{\rm destroyed}}\,,\\
f_{\rm destroyed}^{\rm tides}&=&{N_{\rm tides}^{\rm gal}+N_{\rm tides}^{\rm halo}
\over N_{\rm destroyed}}\,,
\end{eqnarray}

\noindent where 
$N_{\rm destroyed}=N_{\rm merge}+N_{\rm tides}^{\rm gal}+
N_{\rm accr}+N_{\rm tides}^{\rm halo}$. 

By contrast, increasing values of the core radius $r_c$ of the $\beta$ model or increasing the scale radius $r_s$ of the NFW model leads to a reduced number of mergers and a growing number of tide destructions, for instance, at $r_c \geq 400$ kpc or $r_s \geq 300$ kpc tides outnumber mergers.

As per the simulations herein, the destruction of DGs alone cannot explain the observed IC light.  While most of the galaxies destroyed by tides are dwarfs (i.e. run C3), the destruction of a few galaxies of mass $M>10^{11}\msun$ provides more than 60\% of the IC light.  To account for all the IC light, we conclude that some intermediate mass or massive galaxies are getting destroyed by the tidal field of the most massive galaxy.  This is viable as the mass ratios between the most massive galaxy and the destroyed high-mass galaxies are of order 3\,\textendash\,5.  Other possible explanations are: (1) clusters contain many times more DGs than predicted by a Schechter distribution, or (2) DGs have a much lower $\rm{M/L}$ than assumed.  These are both unlikely.

\subsection{Intracluster stars} 

The LF of the observed IC stars in clusters falls well within the predictions of the simulations herein.  The obtained results indicate that the tidal destruction of galaxies (by other galaxies and by the cluster halo) in clusters is sufficient to explain the observed fraction of IC light; this is supported by observational studies, which find a significant IC light component arising from DGs in clusters.

Results from the simulations presented indicate that for the $\beta$ and NFW  models, $f_{\rm ICS}$ increases with the mass of the cluster halo, which is consistent with studies in the literature \cite{lin04,2004ApJ...607L..83M}.

\section{Conclusion} 

\subsubsection{Areas to Improve}
 Including the dynamic friction caused by the halo would result in galaxies falling into the cluster center earlier in the simulation.   We believe starting the interactions (and destructions) earlier would not have a significant effect on the final results;  we believe also that the method used is sound, although generation of the initial conditions involves some tunable parameters.  
 
 Three possible areas of improvement in methodology are  in the treatment of galaxy harassment, the simplified approach with which  galactic encounters are treated (specifically the energy dissipation thereby), and our idealization of an isolated cluster in equilibrium  (although we believe the cluster can be treated as isolated after the epoch of its major mergers).

\subsubsection{Results}
The approach used enabled us to perform a large number of simulations and  cover a large parameter space, while obtaining statistically significant results.  In a simulation using a large number of particles to represent each galaxy it  would be more difficult to establish the relative importance of mergers versus tidal disruptions.

The results and conclusions are as follows. 

\begin{itemize}
\item  The destruction of DGs by mergers dominates 
destruction by tides.  The simulations presented in the foregoing show a dependence on the scale/core radius. 

\item  The destruction of 
galaxies by the tidal field of other galaxies 
and by the cluster halo is sufficient to account for the observed fraction of IC light in galaxy clusters.  There is an increase of $f_{\rm ICS}$ with the mass of the cluster halo, as also supported by the literature.  Our estimate of the IC light fraction is fairly robust.

\item  In clusters simulating the NFW model, the large number of cluster-halo tidal destructions dominates the mass fraction.  In discussing this point, we noted a possible solution to the cusp crisis of DM halos:
the central cuspy region of the cluster DM halo could have inelastic encounters with the member galaxies; this could have the effect of injecting energy into the halo and erasing the cusp. 

\item  In the simulations presented above, the presence of a cD galaxy increases occurrences of accretion, decreases cluster-halo tidal destructions, and reduces the IC stars luminosity fraction. This is opposite to the trend found in observations (where $f_{\rm ICS}$ is higher in cD clusters). 

\item The destruction of high-mass galaxies  ($M>10^{11}\msun$) is required, as dwarfs alone do not contain enough stars to account for the observed IC light, even if they were all destroyed.  A few high-mass galaxies thus must also be destroyed, although the vast majority of galaxies destroyed by tides are DGs. 

\end{itemize}

\chapter{The Cosmological Volume}

\section{Method}

The cosmological volume part of the simulation refined the crude assumption of an isolated cluster in equilibrium.   A standard particle-particle/particle-mesh (P3M) algorithm  \cite{ppp3m} was used with comoving coordinates that expand along with the Hubble flow and facilitate calculations (see Appendix \ref{appcoords}).  Particles representing DGs were created in high-density regions.  These particles were allowed to gravitate together and merge to form more massive galaxies that in turn gravitated together to form clusters.  This required the DG mass to be significantly larger than that of the P3M particles.  In this way the dynamical range of the simulation allowed us to model a cosmological volume of the Universe while resolving DGs.  The simulation of this second half of the project used  the same initial conditions as previous P3M runs.  While working on the isolated cluster I carried out one such run to generate experimental initial conditions.  As in the first part of the project, one module at a time was added to the P3M code in order to conduct tests.

After a calibrating run the parameters of the simulation runs were modified to meet time 
and computational constraints (Appendix \ref{appvol}).  It was desired that the galaxies be significantly more massive than the DM particles. In the first part of the simulation $M_{\rm{min}}=1 \ee{9}\ \msun$ was used (the upper mass range of DGs as defined in section 1.1).  With this value of  $M_{\rm{min}}$, DGs would be only 7 times more massive than DM particles, which was not considered enough.  For this second part of the project then, the value used was  $M_{\rm{min}}=2 \ee{9}\ \msun$.  This value was 14 times that of the mass of the DM particles, which was deemed sufficient.  In order to save time, tests were carried out in a 40\,Mpc\, box, forming 1 or 2 massive clusters with $256^3$ particles.

\section{Creating the particles}

In order to create the initial conditions, I set an initial run using a 40\,Mpc box and $256^3$ particles with a $512^3$ grid.  This involved  changing settings for volume, number of particles, and grid.  This is effected by modifying the `include' files, the {\tt{zeldo.inc}} and {\tt{nadia.inc}} related to the P3M code.  The {\tt{nadia.inc}} file is changed as per the prescription shown in Table \ref{p3mPresc} according to the desired parameters of each run.

After this I modified the parameters of the run in the {\tt{par\_\,z.dat}} file (see Table \ref{seriesf}).  This file contains  the cosmological parameters governing the simulation as well as the run-identifier consisting of three sequential characters that vary according to the given run.  A working knowledge of UNIX was very helpful to manage and handle large data files such as the initial conditions generated.

To test the algorithm for forming galaxies (and later for the subgrid physics) the initial conditions from the 40\,Mpc run mentioned above were used, with one time-saving modification.  The file was modified so as to stop the run when the density  reached $200 \rho_{\rm{mean}}$.  This was because no galaxy forms between $z\approx25$ and $z\approx9$.  The density  reached $200 \rho_{\rm{mean}}$ at $z=8.79$.  All parameters produced for the galaxy particles and the DM particles at this redshift were saved in a file to be used in subsequent tests.

The specific prescription to create DGs {($M=M_{\min}$}) was discussed a few times: it was decided to create galaxies in high-density regions and reduce the mass of neighboring DM particles. In this way, the number of particles (DM + DG) would increase, but the total mass would remain constant.  When a galaxy is created, the code reduces the mass of the neighboring DM particles to conserve mass. The velocity of the created galaxy is also adjusted in order to conserve momentum. A radius $s_i$ is assigned to each galaxy, similarly to what was done in the first part of the project.  A check was implemented to ensure that a galaxy was not created inside the radius of an existing galaxy.  As we had not kept to a simple spherical background cluster with a particular density profile, it was necessary to implement a special way to handle tidal destruction by the background matter.  More specifics will be available in an upcoming article, concerning the cosmological volume part of the simulation (Brito et al. 2009). 
 
Because in the adopted prescription for the creation of galaxies the DM particles and galaxies enter the gravity calculation  in exactly the same way, it was decided, in order to keep track of the galaxy particles, to expand the arrays handling the DM particles and to add the galaxies at the end of these arrays.  

\clearpage

\begin{deluxetable}{ccccccccccccc}
\tabletypesize{\scriptsize}
\tablecaption{{Guidelines file \tt{nadia.inc}}}
\tablewidth{0pt}
\tablehead{
\colhead{NP3} & \colhead{$NP=NP3^3$} 
& \colhead{\emph{N}} &
\colhead{NC} & \colhead{CPU on purplehaze} 
}
\startdata
32 & 32768 & 64 & 23 & less than1 hour \cr
64 & 262144 & 128 & 47 & a few hours \cr
128 & 2097152 & 256 & 94 & a few days \cr
256 & 16777216 & 512 & 189 & a few weeks \cr
512 & 134217728 & 1024 & 379 & several months \cr
\enddata
\label{p3mPresc}
\flushleft 
\end{deluxetable}

\begin{deluxetable}{ccc}
\tabletypesize{\scriptsize}
\tablecaption{Parameter file \tt{par\_\,z}}
\tablewidth{0pt}
\tablehead{
\colhead{value in par\_\,z file} & \colhead{label in par\_\,z file} 
& \colhead{label description (not part of par\_\,z file)}
}
\startdata
Ab1 & ID code & run-identifier  \cr
 0.27 & $\Omega_0$ & $\Omega_0$  \cr
0.04444 & $\Omega_{\rm b}$ & $\Omega_{\rm b}$  \cr
 0.73 & $\lambda_0$ & $\lambda_0$  \cr
71.  & $H_0$ & Hubble constant  \cr
 0.93 & n & Slope of the primordial power spectrum  \cr
  2.725 & T\_CMB & Cosmic microwave background temperature  \cr
 24. & z\_init & initial redshift (\emph{z})  \cr
 30. & Box size & Size of the cosmological volume to be used  \cr
 73661519 & Seed & random seed  \cr
 4. & Cloud Radius & Used in Particle-Mesh
  algorithm\footnotemark \footnotetext{This helps determines how the density is calculated on the grid from the location of particles.} \cr
 \enddata
\label{seriesf}
\flushleft 
\end{deluxetable}

\clearpage

I therefore modified the P3M code 
as follows:

\noindent 1) In order to make the arrays bigger, to accommodate the galaxies, I modified the file {\tt{nadia.inc}} by adding the line: {\tt{parameter (ntotmax$=1.2*{\tt{np}}$)}}.  Note that {\tt{np}} is the number of DM particles.

\noindent2) I replaced the dimension {\tt{np}} of arrays in the code by {\tt{ntotmax}}, as necessary.

\noindent3) I then added to the main code the common block: {\tt{common /galaxies/ ngal}}.

\noindent4) Next, I searched the code for subroutines that looped over all particles and added the common block at the beginning of each subroutine (tantamount to initializing the variable).  I also replaced {\tt{np}} by {\tt{np+ngal}} in each such subroutine, for instance, in the subroutine {\tt{pppm}}.  I repeated this process in the main module.  One exception to these changes is that in the subroutine {\tt{assmass}}, the loop was split into two.  One loop handles the DM and the other handles the galaxies (loops 35 and 135 in the main code).

\noindent5) Additionally, to permit breaking the runs into portions,\ it was necessary to\ specify in the code the initial\ number of galaxies ({\tt{ngal}}).  This\ number was\ specified by adding\ to the {\tt{param.dat}} file\ the line: {\tt{0     number of galaxies}}.  This line could be read by the code after adding, where necessary, the line {\tt{read(1,*) ngal}}.

As a result of these changes a galaxy in the simulations herein can have three possible states: passive,  active, and inexistent.  Galaxies are active when they are created. Then one of two things can happen:

\noindent1) An active galaxy can be destroyed by tides: it becomes passive.

\noindent2) Two active galaxies can merge: one becomes inexistent, the other  remains active and acquires the total mass and momentum of the two galaxies (a tidal destruction followed by accretion is treated as  a merger).
        
\noindent In loops like {\tt{do \  $\tt i=1,{\tt{np+ngal}}$}}   \   or \  {\tt{do}}  \   ${\tt{i}}={\tt{np}}+1$, ${\tt{np}}+{\tt{ngal}}$ there were two possibilities for keeping track of the galaxies after the interaction:  in the P3M code, the inexistent galaxies are excluded but the active and passive galaxies are included; or, in the subgrid physics, the inexistent and passive  galaxies are excluded and only the active galaxies are included.  In this way the code keeps track of the fragments of a galaxy that has been destroyed by tides.  At the end of the simulation clusters are identified and the passive particles in each cluster are examined to find what proportion of IC stars they contain.

After further changes to the code I carried out another test run, changing the final redshift in the {\tt{param.dat}} file.  I plotted relevant results of this run including the {\tt{GAA\_create}} file (contains information about the galaxies being created), the {\tt{fort.91}} file (contains the total mass at every step), and the {\tt{GAA\_cpu}} file.  Soon, the bulk of the galaxy creation module had been tested  and the module was completed.

Next the number of galaxies being produced was adjusted and the modules for merger, accretion, and harassment were tuned.

\noindent$\rm \bf{Galaxy \ creation \ issues}$

A galaxy is created when the density of DM particles at a given cell exceeds a threshold value ($ \rho > 200 \rho_{\rm{mean}}$).   The cells in the grid with $\rm \rho > 200 \rho_{\rm{mean}}$ all have an equal probability of a galaxy being created at that point.   If a cell has a density significantly larger  then $200 \rho_{\rm{mean}}$, it is possible that two galaxies are created there (i.e. if after a galaxy has been created there, but the density is still significantly greater than $200 \rho_{\rm{mean}}$, a second galaxy can be created there at the next timestep).  For conservation of mass, once a galaxy has been created at a given point, surrounding DM particles (around the center of mass) have their mass lowered by the code.  The velocity of neighboring DM particles is also adjusted in order to conserve momentum.

We encountered a problem that the DM particles surrounding cells where galaxies were created had their masses reduced into negative numbers.  To understand this problem, we compared the redshift at which particles started to have negative masses with, for example, the time at which mergers occurred, and with the total mass of galaxies created by a given redshift.  This problem of negative mass particles persisted as successive modules were activated.  Various alternatives were explored in successive runs, for example, varying $bias$.  The quantity $bias$ was introduced in the P3M code to control the probability that a galaxy particle is created at a given cell.  Peak cells are ascribed a certain probability and a galaxy is created, or not, depending on that probability.  The quantity $bias$ has a factor adjustable according to time elapsed in the simulation.  Initially $bias$ was set at $6.0 \ee{3}$, according to the elapsed time in computational units (1 time unit $ = 7.7646\ee{4} {\rm \ Myr}).$  This value of $bias$ makes the cell probability $\sim 0.5$ at $z\approx 9$.

The above problem was explained by the fact that galaxies were being created repeatedly at, or near the same points.  When this occurred, the neighboring particles had their mass decreased successively, causing them to have negative masses at times.  On subsequent test runs, in order to determine how to constrain the value of $bias$, the total mass in the galaxies created was plotted versus the redshift from the file {\tt{GAA\_gal\_Num\_Mass}}.

The problem was finally corrected by changing the probability weighting.  In the code this was done by replacing the line\ {\tt{weight(nbnb)}}$\,\,= $\,\,{\tt{exp(-0.5*dr2/w2)}} by the line {\tt{weight(nbnb)}}$\,\,=\,\,${\tt{pmass(m)*exp(-0.5*dr2/w2)}}.

During the efforts to solve the negative mass problem, and in order to select the value to be used for $\Omega_{\rm gal}$, we experimented with various $bias$ factors by running several test runs up to $z=6$.  I plotted the total mass in galaxies versus redshift in order to determine the appropriate $bias$ for creating a given number of galaxies.  \noindent To obtain these plots, I used an IDL code that plotted the desired quantities from the {\tt{GAA\_gal\_Num\_Mass}} file related to the P3M code.

\begin{figure} [htp]
	\centering
	\begin {center}	
		
				\includegraphics[bb=00 00 605 640,clip=true,viewport=17 195 550 619,
                  scale=.70, angle=0]{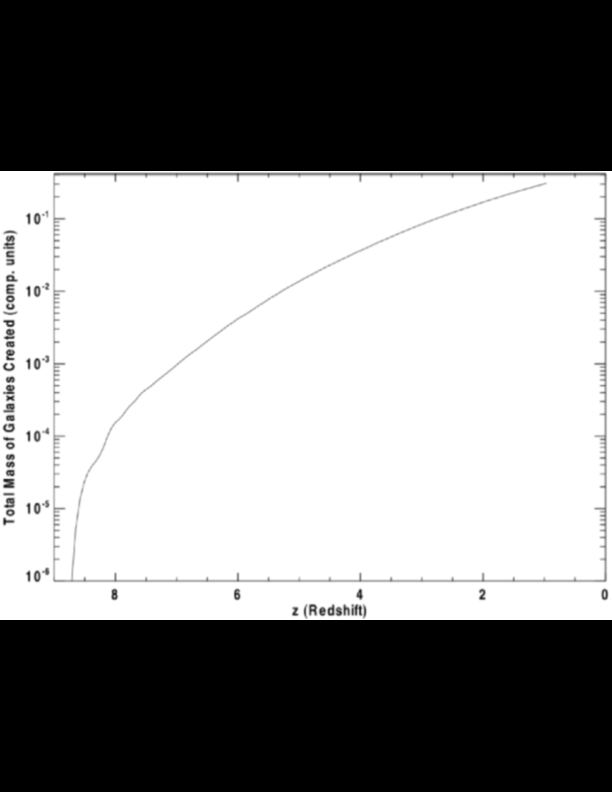}

	\caption[Total mass at a given redshift]{\textbf{Total mass at a given redshift.}  \emph{This represents part of the experimentation carried out to determine the desired value of $bias$.  In this figure $bias=0$.}}
	\label{1bias}
	\end {center}
\end{figure}

\begin{figure} [!htp]
	\centering
	\begin {center}
		
		\includegraphics[bb=00 00 605 640,clip=true,viewport=17 05 600 787,
                  scale=.70, angle=0]{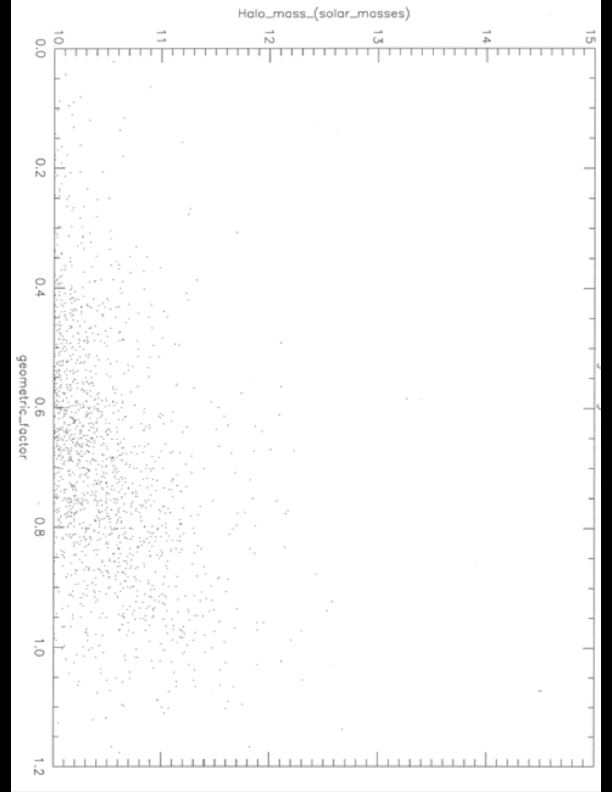}
	\vskip 20pt \caption[Halo mass versus geometrical factor]{\textbf{Halo mass versus geometrical factor.}  \emph{This plot was generated to determine if the geometrical factor used could reproduce a reasonable halo mass.  The linking-length (0.171 in this plot), is important when determining the particles that belong to the halo.}}	\label{linkl}
	\end {center}
\end{figure}

For these various runs we compared the amount of mass being created in the galaxies by plotting versus redshift for different values of $bias$.   A run was executed, commenting the $bias$ factor out,  to determine the maximum mass that could be produced (Fig.~\ref{1bias}).  The value of $bias$ was then adjusted such that the mass ending up in galaxies would total $\Omega_{\rm gal}\approx0.03$.  This test was repeated with this value of $bias$.  The target was to have  about 10\% of the final mass in galaxies, and the rest in P3M particles.  In order to monitor the test runs it was frequently necessary to make plots to compare characteristics of the tests and the simulations; for example Figure \ref{1bias}, with a $bias$ value of 0, served to gauge if enough matter was being created, whereas Figure \ref{linkl} served to monitor the distribution of the matter in the halo.

\section{Test runs for subgrids}

Testing the subgrid physics in the cosmological volume part of the simulation followed a course analogous to that in the isolated cluster.  The subroutines were added one at the time and tested until it was deemed each was working as desired.  We resolved several problems in the course of testing.

\begin{itemize}

\item The subroutine {\tt{accrete}} checks, for each galaxy, if there are DM particles inside the radius of the galaxy.  Particles within that radius are accreted.  The mass, center-of-mass position, and center-of-mass velocity are conserved. The old DM particle still exists, but its mass set to zero.  When the module {\tt{accrete}} was incorporated into the code, I carried out runs and made plots for comparison with previous tests.

The test run including the {\tt{accrete}} subroutine was excessively slow, i.e. it  took 327 steps to reach $z=6$ (29 hours) compared with previous runs without the {\tt{accrete}} subroutine (92 steps to reach $z=6$, duration $<18$ hours).  This problem was solved by changing the timestep subroutine  to exclude from the calculation of the timestep particles whose mass had been set to 0. This increased the length of the timesteps and reduced the run duration.

\item Analogous problems were encountered in the subroutine {\tt{creategal}}: it allowed galaxies that had been destroyed to continue having interactions with other galaxies (their mass had been diminished to zero but the particles had been kept to keep track of the IC medium).  This problem was fixed by excluding the massless particles from those loops after the merger had been processed.  

\item The same problem as above was encountered in the tidal destruction subroutine.

\item There appeared to be runaway increase in sizes of galaxies.  At times galaxies attained  almost 1/10th of the total mass inside the box.  Subsequent multiple harassments  at the same timestep caused the galaxy to experience an uncontrolled increase in size, beyond the size of the box.  A size cutoff was stipulated for the galaxies in the simulation by incorporating a cutoff size for the galaxies in the parameter $sizMAXcgs$ initialized in the subroutine {\tt{ComovToPhysical}}.
\end{itemize}

As a side effect of excluding the massless particles, there were fewer interactions in the test runs.  In contrast to the isolated cluster, where galaxies grew by mergers only, in this, the cosmological volume part of the simulation, galaxies could grow by accretion.  This resulted in a lower number of mergers.  A test run was conducted after the exclusion of the massless particles.  The test run created 9682 galaxies with a total mass of 0.3079 (including those tidally disrupted and those going to IC light).

A histogram of the mass distribution of galaxies in the run including accretion showed that the total mass in the galaxies followed the prescription used herein but that the mass distribution did not follow the Schechter distribution (therefore was not correct according to observations).  Indeed, in the obtained distribution there were very few DGs.  This problem was addressed by fixing a threshold lower than $200 \rho_{\rm{mean}}$ in the galaxy formation criterion.

The test runs had begun, initially, in  a $\rm 40\,Mpc$ box ($M_{box} \approx 2.4\ee{15} \ M_\odot$).  While these runs yielded a reasonable cluster mass, it was necessary to redo several test runs, because of the change in galaxy formation criterion to a threshold lower than $200 \rho_{\rm{mean}}$  (as a fix for one of the bugs).  It was also considered that in view of time pressures future test runs would be effected in a $20\rm{\,Mpc}$ box with \ $128^3$  particles.

The new test runs thus used a 20\,Mpc box (total mass $M_{box} = 2.95\ee{14} M_\odot$).  A problem ensued: the simulation produced only a very small mass of IC stars.   The vast majority of the tidal disruptions led to accretions, with little material going to the IC medium.  It was thought that, as the IC light fraction increased with increasing cluster halo mass, this problem would disappear  when simulating a much bigger box.  Another possible explanation was that, with the current small volume used in the simulation, the galaxies never encountered the gravitational potential of any massive cluster (since the clusters forming in a 20\,Mpc box are much less massive) and their resulting velocities were thus smaller.

It was also thought that some galaxies had simply been created very close to each other, their gravitational potential energy thus being higher than their kinetic energy, leading to quick accretions.  It was considered that if the merger condition were loosened by checking for mergers whenever $r_{ij}=S_i+S_j$, some of the accretions might be counted as mergers, thus solving the problem.

A test run was effected at this point, with the parameters modified as detailed above and including the subgrid physics.  The time for this full test run was $\sim$ two weeks.

\begin{figure} [!htp]
\begin{center}

	\includegraphics[bb=00 00 605 640,clip=true,viewport=12 63 600 730,
                  scale=.79, angle=0]{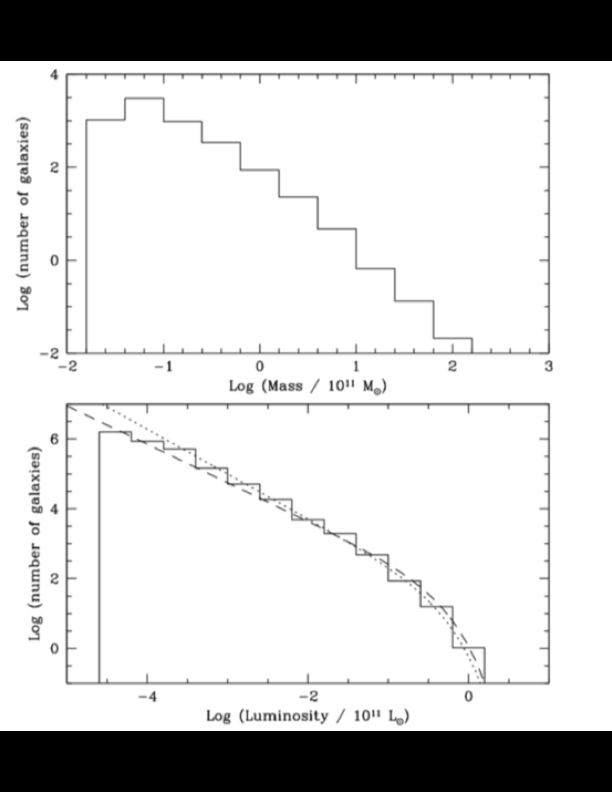}

\vskip-10pt
\caption[Schechter function reproduced]{\textbf{Schechter function reproduced.}  \emph{Top panel: mass distribution of the galaxies at $z=0$.
Bottom panel: luminosity distribution of the galaxies at $z=0$;
(dotted curve: Schechter luminosity function with $\alpha_{\rm start}=-1.28$;
dashed curve: Schechter luminosity function with $\alpha_{\rm start}=-1.10)$.}\label{folschech}}
\label{followschechter}
\end{center}
\end{figure}

\begin{center}
 \begin{figure} [!htp]
 	\hskip-50pt
	\centering
		\includegraphics[angle=270, width=1.0\textwidth]{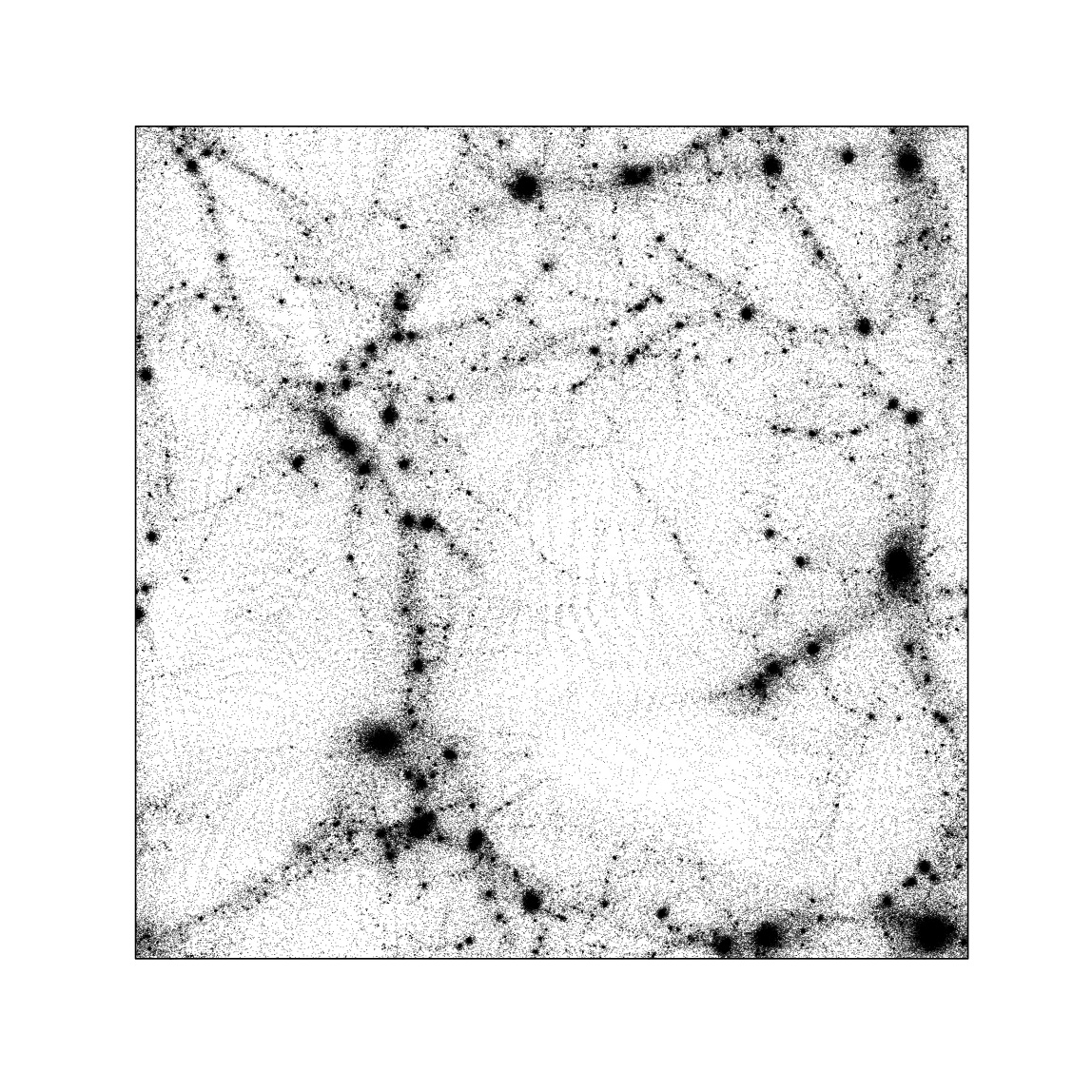}
	 	\vskip-10pt \caption[Cosmological volume slice]{\textbf{Cosmological volume slice.}  \emph{Final state at $z=0$ of the simulation that produced the luminosity function shown in Figure \ \ref{followschechter}.  This run used a 20\,Mpc box; the plot shows the location of the P3M particles in a slice of thickness 4\,Mpc.}}
	\label{cosmoics}
 \end{figure}
\end{center}

As the tests progressed the mass distribution plots revealed that the simulation was beginning to produce realistic results.  Figure \ref{followschechter} was used to monitor the shape of the corresponding luminosity function, and Figure \ref{cosmoics} was used as another way to monitor the distribution of matter and voids.

\section{Conclusions}

The cosmological part of the simulation presented herein is at a very exciting point at present. The modules for subgrid physics are working in a realistic way.  The latest test run produces a Schechter-like distribution (Fig.~\ref{folschech}).  This test run includes the galaxy creation module as well as the mergers module.  A slice of the distribution of particles in the 20\,Mpc volume is shown in Figure~\ref{cosmoics}.  This shows a slice of thickness corresponding to 4\,Mpc of a test run in a 20\,Mpc box.  There is an ongoing problem: the remaining module, for tidal disruption by the background halo, does not work as desired in the test runs.  As mentioned in the foregoing, there is currently not enough matter being imparted to the IC medium.\\  The reason for this could be the small size of the test box used and the correspondingly small amount of matter in this box.  It is thought that there is a number of accretions that would be counted as mergers, were the galaxies moving in a more massive system  (the merger criterion was satisfied too easily in the reduced volume).  Were the galaxies experiencing interactions, i.e. encounters, or background halo tides, they would have larger kinetic energy (hence larger velocities).  Tests while correcting this problem are expected to be time-consuming beyond the time allotted for a master's degree research project, hence I obtained permission to present the results to-date.  Further progress will be presented in an article (Brito et al. 2009).  The Schechter distribution reproduced from tests until now, combined with the success of the merging module, and previous encouraging results from the isolated cluster (agreement with accepted theory) make us optimistic that this problem will shortly be resolved.  We will then be able to launch the main 80\,Mpc simulation, expected to take $\sim$1 month to complete.  Ideally, the analysis of that simulation will show that the results from the isolated cluster carry over to an actual cosmological volume where non-spherical clusters have evolved under gravity.

The next step is to `turn on the tides' (include the module that keeps track of tidal forces by the halo) and carry out test runs that will tell us if we are ready to expand the test volume to the full 80\,Mpc cosmological volume.  These test runs were originally to be carried out in a 20\,Mpc volume, but will instead be carried out in a volume large enough that the total mass in the box will be at least comparable to the mass of a cluster.  In order for the numbers obtained to be significant statistically, the final run must be carried out in a volume large enough to form several clusters ($\sim$100\,Mpc). 

The results for the isolated cluster simulation (part 1 of this project) were in agreement with the fraction of IC stars currently thought to be due to tidal disruption.  If this cosmological volume simulation (part 2 of this project) continues to reproduce numbers comparable to the accepted theory, and if the distributions of $\rm{M/L}$ continue to reproduce a Schechter function it will be strong support for the validity and significance of the method used.

\appendix
\chapter{Article 1: The Fate of Dwarf Galaxies in Clusters and \\ the Origin of Intracluster Stars} \label{appart}  
This article was accepted in March, 2009, for publication in the  Journal of Astrophysics and Astronomy, Volume 30, Number 1, pp. 1-36.  It is also available at arXiv.org: arXiv:0707.1533

The text of the article was written by Dr. Martel, while the subroutines added to the code that modeled the simulations were written by Dr. Barai.  My contribution is as detailed in Chapter 2 of this thesis.

\chapter{Interpolation tables} \label{appinterp}

This appendix constitutes the latest version of the procedure that calculates the interpolation tables from which the mass of the cluster halo is obtained for use in the simulations.  This is a routine written in FORTRAN; the asterisk at the beginning of lines denote comments.

\begin{verbatim}

* This program builds interpolation tables for the density
* profile of haloes. 
*
* To calculate values for a different density profile simply
* 'comment out' the current profile and remove the comment marker
*  for the desired profile.  The alternative profiles are located towards the end
* of the user-defined function at the end of this procedure (maketable).
*
* This procedure comprises three parts.  The first part creates the table
* and calculates table entries using an integral calculated in the second part.
* The second part uses a user-defined function created in the third part.
*

      program maketable
*
* Code written by Hugo Martel, modified by William Brito, 2007.
*
      implicit double precision (a-h,o-z)

      external g

      parameter (pi=3.1415926535)

      npt=10000
      smax=20.

      ds=smax/npt

* Open output file

      open(unit=1,file='profile1.out',status='unknown')

 1001 format(5x,1pe13.6)

      a=0

      write(1,1001) smax
      do i=0,npt
           s=float(i)*ds
           call integ(a,s,g,area)
           f=4.*pi*area
           write(1,1001) f
      enddo

      close(unit=1)
      stop
      end

*====================================================================
      subroutine integ(a,b,f,area)
*
* This subroutine computes integrals using Simpson's rule.
**====================================================================
      implicit double precision (a-h,o-z)

      data tol /1.e-05/

      if(a.eq.b) then
           area=0.
           return
      endif

      areaold=0.
      sumend=F(a)+F(b)
      sumeven=0.
      sumodd=0.

      do n=1,25
           i=2.**n
           h=(b-a)/dfloat(i)
           sumodd=0.
           do j=1,i-1,2
                c=a+j*h
                sumodd=sumodd+F(c)
           enddo
           area=(sumend+4.*sumodd+2.*sumeven)*h/3.
           if(dabs((area-areaold)/area).lt.tol) return
           sumeven=sumeven+sumodd
           areaold=area
      enddo

      write(6,1000)
 1000 format(/5x,'Error, no convergence.')
      stop

      end




*====================================================================
      function G(x)
*
* This function calculates the integrand, and passes it to 
* the procedure integ to integrate.
*
* the integrand consists of the product of a density function (which is
* a function of x) and x squared, so as to integrate over a volume
* and obtain the mass.
**====================================================================

* G(x)=f(x)*x*x
*
*
      implicit double precision (a-z)

      if(x.eq.0.) then
         g=0

      else

* NFW 
* To calculate values for a different density profile simply
* 'comment out' the current profile and remove 
* the comment marker for the desired profile.
*
*      g=x*x/(x*(1.+x)**2)

* Hernquist
*
*      g=x*x/(x*(1.+x)**3)
*
* xi model (using the best fit values xi=1.4 and a=.33*r200
* from Biviano and Gerardias. There is another value of xi also 
* given there, check it out now!)  
* Note xi=1 corresponds to NFW profile.
*
*      g=((r/a)**-1.4)*((1+r/a)**(1.4-3))
*
* King profile (Beta=1)
*
      g=x*x/(1.+x**2)**(3/2)
*

* The beta model (in Biviano and Gerardi this is credited to 
* Cavaliere and Fusco-Femiano (1978) but
* it is worthwhile to check the original reference!) using
* a best-fit value of .8 from Biviano and Gerardi as well as several of 
* Piffaretti and Kaastra's best fit values for clusters listed on their Table 2.
*
*      g=x*x/(1.+x**2)**((3/2))*.8
*      g=x*x/(1.+x**2)**((3/2))*.46
*      g=x*x/(1.+x**2)**((3/2))*.33
*      g=x*x/(1.+x**2)**((3/2))*.59
*      g=x*x/(1.+x**2)**((3/2))*.32
*      g=x*x/(1.+x**2)**((3/2))*.53
*      g=x*x/(1.+x**2)**((3/2))*.49
*      g=x*x/(1.+x**2)**((3/2))*.45
*      g=x*x/(1.+x**2)**((3/2))*.50
*      g=x*x/(1.+x**2)**((3/2))*.52
*      g=x*x/(1.+x**2)**((3/2))*.42
*      g=x*x/(1.+x**2)**((3/2))*.55
*      g=x*x/(1.+x**2)**((3/2))*.38
*      g=x*x/(1.+x**2)**((3/2))*.62

      endif

      return

      end


\end{verbatim}

\chapter{Coordinate conversion} \label{appcoords}

The particle-particle/particle-mesh (P3M) code used in this project uses comoving coordinates, not physical coordinates; the computational volume therefore expands with time.  Although this means that the simulation box is expanding, the galaxies themselves are not expanding. The box expands along with the Universe, so that the mass  inside the box remains constant.  The box gets larger and larger relative to the galaxies. If coordinates are used in which the box had a fixed size of 1, then in those coordinates the galaxies would shrink with time.  In physical units the box size and the initial size of galaxies scale like $1/(1+z)$, so in computational units, they are both constant.  The box expands with time, but the galaxies do not.  In computational units, the box is fixed and the galaxies shrink.

The P3M code used utilizes a special form of comoving coordinates, denoted by a tilde, called $\it{supercomoving}$ $\it{variables}$ \cite{1998MNRAS.297..467M}.  
To convert the peculiar velocities to actual velocities it is necessary to convert the coordinates before calculating quantities such as the kinetic energy.

\noindent In supercomoving variables there is a precise normalization for the scale factor.  The  normalization depends on the particular cosmological model. For the models considered here (where $\Omega_0$ is the density parameter, $a$ is the scale factor, $\lambda_0$ is the cosmological constant, and the subscript `0' denotes the values at the present time) the solution of the Friedmann equation and the present value of the scale factor are, as per Martel and Matzner~\cite*{supercom}:

\noindent 1. Einstein de Sitter model: ($\Omega_0 = 1, \ \lambda_0=0$): 

\begin{equation}
	\normalsize {a=\tilde{t}}^{-2}, \ \normalsize {a}_0=\normalsize {1}.
\end{equation}

\noindent 2. Open models ($\Omega_0 < 1, \ \lambda_0=0$):
\begin{equation}
	\normalsize {a}={\normalsize {1}\over \normalsize {\tilde t}^{2}-1}, \ \normalsize {a}_0={\normalsize {1-\Omega}_0\over \normalsize {\Omega}_0}.
\end{equation}

\noindent 3. Flat models with non-zero cosmological constant ($\Omega_0 + \lambda_0 = 1):$
\begin{equation}
	\label{flatLambdmods}
	\tilde{t} ={1\over2}\int_1^a \frac{dy}{y^{3/2}(1+y^3)^{1/2}}, \ a_0 = \left( \frac{\lambda_0}{\Omega_0} \right)^{1/3}.
\end{equation}

The solutions for $a(\tilde{t})$ do not depend explicitly upon the cosmological parameters (these are absorbed in the definition of $a_0$).  Therefore, for all the models included in the database (the database as in Martel \& Matzner 1999), there are only three different solutions of the Friedmann equation.

The positions $s$ and velocities $\bf{u}$ in physical units are given by:

\begin{equation}
\label{spacord}
\emph{r} = \frac{L_{\mathrm{box}} \bf \tilde{\emph{r}} }{1+z},
\end{equation}

\begin{equation} 
\label{}
 \mathbf{u} = L_{\mathrm{box} }  \left[\frac{ H(z)\mathbf{ \tilde{\emph{r}} }} {1+z} + \frac{\Omega_0^{1/2}H_0(1+z)\tilde{\mathbf{v}}}{2a_0^{1/2}}\right],
\end{equation}

\noindent where $H(z)$ is the Hubble constant as a function of redshift ($H_0$ denoting its value at the present epoch).  The size of the box being considered is $L_{\mathrm{box} }$ and $\bf{v}$ is the peculiar velocity.

\chapter{Box Volume and Particle Size} \label{appvol}

In the run used as guide to determine the timeframe for the project, the number of particles used in the cosmological volume (120\,Mpc box) was $128^3$.  The mass per particle this implied\footnotemark \footnotetext{In computational units the mass of the box is 1, the dimension of the box is $1\times1\times1$, and thus the mean density is 1. Since the mass is 1, for $NP$ equal-mass particles, the mass per  particle is $1/NP$ (as in the {\tt{zeldo.inc}} code).  In physical units the box is $L_{{\rm{box}}} \times L_{{\rm{box}}} \times L_{{\rm{box}}}$.  The  volume is thus $L_{{\rm{box}}}^3$.  The mass inside the box is the volume times the mean density inside the box. (The box  represents a cosmological volume so its mean density is equal to the mean density of the Universe $\rho_{\rm{mean}}$.) Thus:  $M_{{box}}  =  \rho_{\rm{mean}} \times L_{{\rm{box}}}^3  =  \Omega_0 \times \rho_{\rm crit} \times L_{{\rm{box}}}^3  =  3 \times H_0^2 \times \Omega_0 \times L_{{\rm{box}}}^3 \times  1/ 8 \pi \rm G$.  Note that as usual  $\rm G$ is the gravitational constant, $H_0$ the Hubble constant, and $\Omega_0$ the density parameter.  Dividing the mass of the box by $128^3$ gives the mass per particle.  } was $\sim 3
\ee{10}\ \msun$\,\textemdash\,approximately 30 times the minimum galaxy mass used in the isolated cluster simulations  ($M_{\rm{min}} = 1\ee{9} \ \msun$).

To avoid carrying out a simulation that would last several months we could not simply increase the number of particles arbitrarily.  At that time a final run had been envisioned using $512^3$ particles.  This would reduce the mass to $4.86\ee{8}\ \msun$ (smaller than $M_{\rm{min}}$ by a factor of 2).

Using an 80\,Mpc box the particle mass would be further reduced to $1.44 \ee{8}  \msun$  (about $1/7  \ M_{\rm{min}})$, which would allow increasing ${M}_{\rm{min}}$ to $2 \ee{9}\ \msun$ (a factor of 2).  This $M_{\rm{min}}$ 14 times larger than the particle mass used in the isolated cluster simulations was deemed to be a good choice.  The estimated time for such a run to complete was $\sim $1 month.  With these modifications it was decided to `tune' the code running tests with a box size of 40\,Mpc; the 1 or 2 massive clusters that would be produced were deemed adequate. In tests, $256^3$ particles would be used to reduce running time.  For this volume, the mass scaling yielded $256^3$ particles.  These tests would require only a few days and would be carried out as the code advanced.  For example, a test run was to be carried out once the addition was effected of the code that accounts for tidal force by the background matter (the cluster was no longer restricted to being spherical).  These tests would serve, for example, to ensure that there were enough galaxies being formed in the simulation and to ascertain that the luminosities of galaxies created,  actually followed a Schechter distribution (Fig.~\ref{followschechter}).

\bibliographystyle{astron}

\bibliography{thesis_postCaramelization}

\addcontentsline{toc}{chapter}{\bf{E\,\,\,Bibliography}}
\end{document}